\documentclass[apj,numberedappendix]{emulateapj}
\slugcomment{Accepted to ApJ: June 28, 2010}
\usepackage{amsmath}
\def\gtorder{\mathrel{\raise.3ex\hbox{$>$}\mkern-14mu
                \lower0.6ex\hbox{$\sim$}}}
\def\ltorder{\mathrel{\raise.3ex\hbox{$<$}\mkern-14mu
                \lower0.6ex\hbox{$\sim$}}}

\shorttitle{Equilibrium Figures of Binary KBOs}
\shortauthors{Gnat \& Sari}

\begin{document}
\title{Equilibrium Configurations of Synchronous Binaries: Numerical Solutions and Application to Kuiper-Belt Binary 2001 QG$_{\rm 298}$}
\vspace{1cm}
\author{Orly Gnat\altaffilmark{1,2} and Re'em Sari\altaffilmark{3,1}}
\altaffiltext{1}{Theoretical Astrophysics, California Institute of Technology, 
        MC 350-17, Pasadena, CA 91125.}
\altaffiltext{2}{Chandra Fellow}
\altaffiltext{3}{Racah Institute of Physics, Hebrew University, Jerusalem 91904, Israel}
\email{orlyg@tapir.caltech.edu}

\begin{abstract}
We present numerical computations of the equilibrium configurations 
of tidally-locked homogeneous binaries, rotating in circular orbits.
Unlike the classical Roche approximations, we self-consistently account for
the tidal and rotational deformations of both components, and relax the 
assumptions of ellipsoidal configurations and Keplerian rotation.
We find numerical solutions for mass ratios $q$ between $10^{-3}$ and $1$,
starting at a small angular velocity for which tidal and rotational
deformations are small, and following a sequence of increasing angular
velocities.
Each series terminates at an appropriate ``Roche limit'', above which no 
equilibrium solution can be found. Even though the Roche limit is crossed
before the ``Roche lobe'' is filled, any further increase in the angular
velocity will result in mass-loss. 
For close, comparable-mass binaries, we find that local deviations from 
ellipsoidal forms may be as large as $10-20\%$, and departures from 
Keplerian rotation are significant.
We compute the light curves that arise from our equilibrium configurations, 
assuming their distance is $\gg1$~AU (e.g. in the Kuiper Belt). 
We consider both backscatter (proportional to the projected area) and diffuse
(Lambert) reflections. Backscatter reflection always yields two minima of
equal depths. Diffuse reflection, which is sensitive to the surface 
curvature, generally gives rise to unequal minima. We find detectable 
intensity differences of up to $10\%$ between our light curves and those 
arising from the Roche approximations. Finally, we apply our models to Kuiper
Belt binary 2001~QG$_{298}$, and find a nearly edge-on binary with a mass ratio 
$q=0.93^{+0.07}_{-0.03}$, angular velocity $\omega^2/G\rho=0.333\pm0.001$
(statistical errors only), and pure diffuse reflection.  For the observed 
period of 2001~QG$_{298}$,  these parameters imply a bulk density, 
$\rho=0.72\pm0.04$~g~cm$^{-3}$.

\end{abstract}

\keywords{Kuiper Belt -- minor planets, asteroids -- solar system: general}

\section{Introduction}
\label{intro}

The Kuiper Belt consists of a large number of small objects
in heliocentric orbits beyond Neptune. The existence of the
Kuiper Belt was suggested on theoretical grounds by 
Edgeworth (1943) and Kuiper (1951), but it was not 
until 1992 that the first Kuiper Belt object (KBO) was
detected (Jewitt \& Luu~1993).
To date, more than $1000$ KBOs are known.
They are thought to be relics of the Sun's accretion 
disk, and to hold signatures of the planetary migration 
process. Their physical properties, including their
mass-distribution, compositions, and binary-fraction may
thus hold the key to our understanding of the formation and
evolution of the early solar system. A comprehensive review
is provided by Luu \& Jewitt~(2002).

The properties of Kuiper belt binaries place important constraints
on theories of solar system evolution. In particular, the distributions 
of separations and mass ratios are a unique signature of
the binary formation process. Over the past decade, more than
$50$ resolved Kuiper belt binaries have been discovered and
studied (Noll et al.~2007, 2008). These can be broadly divided into two groups
(Noll et al.~2008). The first consists of small satellites
in orbits about larger KBOs. These systems are believed to have formed
through a large collision, followed by tidal evolution, much like the
Earth-Moon and Pluto-Charon systems (Hartmann \& Davis~1975; Cameron \&
Ward~1976; McKinnon~1989). 
The second - more abundant - group, consists of comparable mass, large 
separation binaries. These are inconsistent with
the classical formation mechanisms, and have prompted an abundance of new
theoretical models (Weidenschilling~2002; Goldreich et al.~2002; 
Funato et al.~2004; Astakov et al.~2005; Lee et al.~2007; Schlichting 
\& Sari~2008; Gamboa Su{\'a}rez et al.~2010; Naoz et al.~2010). 
Recently, Sheppard \& Jewitt~(2004) postulated the existence of a 
third group of Kuiper belt binaries. By following the light curves of 
KBOs, they were able to identify an unresolved, comparable-mass 
small-separation binary-candidate, 2001~QB$_{298}$, based on its 
variability, photometric range, and period. Their analysis indicates 
that at least $10-20\%$ of all large KBOs may in fact be unresolved 
close-binary systems.

Our current interpretation of the observed light curves of
KBOs relies on the classical theory of the equilibrium figures
of rotation. 
The equilibrium configurations of rotating fluid bodies
is a classical problem that was first investigated
by Newton in the context of the figure of the Earth.
Newton showed that the slow rotation of the Earth
causes it to become a slightly oblate spheroid.
In 1742, Maclaurin generalized the problem, by relaxing the
assumption of slow rotation. He derived a general relation
between the velocity of rotation and the eccentricity of the
resulting equilibrium spheroid. Maclaurin's relation implies
that for any given angular velocity, two equilibrium 
solutions may be found, with different eccentricities.
Decades later, Jacobi (1834) realized that triaxial
ellipsoids are also possible equilibrium configurations
of homogeneous rotating masses.

The study of possible equilibrium configurations of binary
systems was introduced by Roche (1847). Roche considered
the equilibrium form of a satellite rotating about a rigid
sphere in a circular Keplerian orbit. The satellite is deformed
both by rotation and by the tides exerted by its spherical companion. 
A more accurate approach was later taken by
Darwin (1906), who revisited the problem, allowing for
mutual deformations of both bodies. 
Solutions to Darwin's problem may be found only in the limit
of equal mass ratio ($q=1$) or extreme mass ratio ($q\ll1$).
Both the Roche and Darwin solutions rely on the assumption 
that the equilibrium configurations are triaxial ellipsoids. 
A thorough review, analytic derivation, and stability analysis
of these classical equilibrium configurations can be found
in Chandrasekhar~(1969).

In modern astrophysical research, these classical Roche ellipsoidal
approximations are used in the study of a wide range of 
contact-binaries, including binaries in the solar system. 
Weidenschilling (1980) used equal-mass Darwin models to fit the
observed light curves of asteroids 624~Hektor and 216~Kleopatra.
In 1984, Leone et al. analytically constructed arbitrary mass-ratio Roche
ellipsoidal models. When computing a Roche binary model for an 
arbitrary mass ratio, each component is in turn assumed to be 
spherical while the ellipsoidal configuration of its companion 
is calculated. Leone et al.~(1984) used their theoretical
photometric-range as a function of the angular velocity to 
constrain the mass ratios and densities of observed variable
asteroids. Cellino et al.~(1985) explicitly computed the light curves
arising from the equilibrium Roche models of Leone et al., 
and compared them with an observed sample of asteroid light curves
to constrain their orbital parameters and densities.
Jewitt \& Sheppard~(2002) used the observed period and photometric
range of KBO 20000~Varuna to consider possible (single) Jacobi ellipsoid
and Roche binary models for this object. Takahashi \& Ip~(2004) computed
the light curves arising from Roche ellipsoidal configurations, to confirm 
the nature of the suspected binary KBO 2001~QG$_{298}$. Lacerda \& 
Jewitt~(2007) constructed a library of Jacobi-ellipsoid and Roche-binary
light curves,  and investigated the nature and densities of four variable KBOs
(20000~Varuna, 2003~EL$_{61}$, 2001~QG$_{298}$, and 2000~GN$_{171}$) and
of the Trojan asteroid 624~Hektor. Descamps~(2008) applied Roche binary
models to a sample of variable asteroids.

All of the binary models mentioned above rely on the assumptions of the
Roche approximation, namely that each component is deformed by the tides of a 
{\it spherical} companion, and that the resulting configurations are 
ellipsoids rotating in Keplerian orbits. In this paper we present
new numerical computations of the equilibrium configurations of tidally 
locked homogeneous binaries, orbiting in circular orbits. We self-consistently
take into account the tidal and rotational deformations of {\it both}
components, and relax the assumptions of ellipsoidal configurations
and Keplerian rotation. For comparable-mass small-separation binaries,
departures from ellipsoidal configurations and Keplerian rotation
become significant.

Numerical solutions of the equilibrium configurations
were previously computed for the cases of general mass ratio homogeneous 
binaries (Hachisu \& Eriguchi~1984a) and for equal-mass polytropic binaries 
(Hachisu \& Eriguchi~1984b). These previous works expanded the local 
gravitational potentials in terms of Legendre polynomials, which were used to
approximate the potential on the surface of the two components.
In our models we explicitly compute the local gravitational and rotational
potentials on the surface of the two bodies, and use a Newton-Raphson based
scheme to converge to an equilibrium solution, for which the surface of each
of the components is an equipotential surface.
We then use our numerical
solutions to calculate the light-curves that these equilibrium 
configurations exhibit if placed in the Kuiper Belt, and demonstrate
how these models can be used by fitting the observed light curve of
Kuiper belt binary 2001~QG$_{298}$.

In our computations we make several simplifying assumptions.
First, we assume that the bodies' gravity dominates over their
internal strength, so that they take the forms of rotating 
{\it fluid} binaries appropriate for their angular velocity.
The ratio of the material rigidity to self-gravity determines a 
size scale above which bodies may be considered gravity-dominated.
For rigid monoliths, this size scale is $\sim10^4$~km. 
However, many of the smaller solar-system bodies are expected to be
``rubble-piles'', for which the effective rigidity is reduced
(Goldreich \& Sari~2009; and references therein). 
Rubble piles larger than a few hundred kilometers are likely to be 
gravity dominated. In addition, over the age of the solar system, 
even smaller bodies will gradually take the forms dictated by 
gravity.

Second, we assume that the densities of the two binary components
are equal. While this is likely the case for binaries formed through
collisions, it is not necessarily true for binaries created via
three-body interactions or due to dynamical friction. At the opposite
extreme, when the density ratio is infinite, one of the bodies is
effectively compressed into a point mass while the other takes
a form similar to the classical Roche solution. These two limiting
cases therefore span the range of possible configurations for 
different density-ratios.

Third, we assume that the binaries are tidally locked, so that
their orbital and spin frequencies are equal, and their orbits
circularized. If the orbital and spin velocities differ, the tidal
bulge is carried ahead (or lags behind) the companion. Torques then
act to re-align the system, inducing oscillations about the
equilibrium aligned configuration. Internal dissipation 
damps these oscillations, and results in synchronized circular
orbits. For the fiducial physical properties of large KBOs ($R\sim100$~km,
$\rho\sim2$~g~cm~$^{-3}$, $k_{\rm rubble}\sim10^{-3}$, $Q\sim100$), 
such synchronization is expected to occur on time scales that are
shorter than the age of the solar system
for binaries forming with initial separations $\lesssim 50R$ (Goldreich
\& Sari~2009).

The outline of this paper is as follows.
In section~2, we briefly describe the numerical method that we use
to compute the equilibrium configurations as a function of the mass
ratio and angular velocity. Additional details are provided
in APPENDIX~A.
In Section~3 we discuss the equilibrium figures of rotation. We first use analytic
approximations to clarify the nature of the equilibrium solutions and 
the evolution surrounding the ``Roche limit''.
We then present exact solutions for the equilibrium configurations
using our numerical approach.
In Section~4 we show that our equilibrium configurations are generally 
non-ellipsoidal, and deviate from pure Keplerian rotation. We explain how
the angular momentum depends on the mass ratio, angular velocity
and tidal deformations.
In Section~5 we compute the light curves that arise from the numerical
configurations (the numerical procedure used to compute the light curves
is described in APPENDIX~B).
We explain how the reflection properties of the materials
affect the observed variability, and demonstrate the differences between 
our light curves and those resulting from the classical Roche approximations.
Finally, in Section~6, we use our models to fit the observed properties
of Kuiper Belt binary 2001~QG$_{298}$, and to derive the physical properties
and bulk density of this system.
We summarize in Section~7.

\section{Numerical Method}
\label{method}

We are interested in finding the self-consistent
equilibrium configuration of tidally-locked homogeneous binaries,
rotating in circular orbits. 
We assume that the bodies are strengthless, so that their
equilibrium configurations are determined by the conditions that
the total (gravitational $+$ rotational) potential is constant along 
the surface of each of the bodies (but may differ between the two bodies).

The parameters that determine the equilibrium configurations are
the mass ratio $q\equiv M_2/M_1<1$; the separation between the
components $d$; and the scaled angular velocity, $\omega^2/G\rho$.
Here we treat the separation and angular velocity as two independent
parameters, and solve for the total mass of the system.
For non-spherical configurations, the mass, separation, and angular velocity
are not related through Kepler's law, and our choice of parameters allows
us to conveniently explore the non-Keplerian nature of the solutions.
Our numerical procedure is described in detail in~\ref{equations}. 
We summarize the numerical scheme below.

We parametrize the surface of each body by specifying the
distance from its center to the surface along $N$ preset
angular directions, $R_{i=1...N}$. We associate each of the $N$ surface-points
with the small surface area surrounding it, and with a 
``mass-cone'' stretching from the center to this surface area.
The mass in the $i$'th cone is then $M_i=\Delta\Omega \rho R_i^3/3$,
where $\Delta\Omega$ is the solid angle that the cone occupies. 
We distribute the $N$ points evenly, so that each cone covers
an equal solid angle as viewed from the center of the body.

The potential at a point located on the surface is then 
the sum of the gravitational potential induced by the mass
in all cones of both components, and the rotational 
potential about the common center of mass.
For a model with $N_1$ points sampling the surface of $M_1$, and
$N_2$ points sampling the surface of $M_2$, our goal is 
to solve for the $N_1+N_2$ values of $R_i$ for which
the total potential along the surface of each body is constant.
The value of the potential may differ between the two bodies.

We explicitly compute the gravitational and rotational 
potentials (see~\ref{equations}). Our algorithm solves 
for $N_1+N_2+2$ variables, namely the values of the distances
$R_i$ defining the surfaces of the bodies, and the values
of the  constant potentials along their surfaces, $c_1$ on $M_1$ and
$c_2$ on $M_2$. The equations that we solve are the $N_1+N_2$
conditions of constant potential, and the additional constrains
provided by the given mass ratio $q$, and separation $d$,
\begin{equation}
\scriptsize
\label{F}
\vec{F} \equiv  \left\{
\begin{array}{l}
\sum_{i\in M_1, M_2}V_i(\vec{R}_j) +
V_{\rm rot}(\vec{R}_j) + c_{k}\;\;,\;\;{\rm for~}j\in M_k \\
M_1/M_2-q \\
x_{{\rm CoM},1} - x_{{\rm CoM},2} + d
\end{array}
\right\}
=0\,,
\end{equation}
(see~\ref{equations}), where $x_{{\rm CoM},k}$ is the center of mass
position of $M_k$, along the direction separating the two components.

We use a Newton-Raphson scheme to solve for the variables,
$\vec{x}=(\vec{R}_j, c_1, c_2)$ that satisfy equations~(\ref{F}).
We start with an initial guess for the configurations,
and in each iteration correct the current guess by
a small amount $\vec{dx} = -\vec{F} \times {\bf J}^{-1}$,
where ${\bf J}$ is the Jacobian derivatives matrix.
We compute the Jacobian analytically, and iterate the
Newton-Raphson scheme until the solution has converged
so that $\vec{F}=0$ to within some numerical
threshold, and the correction $\vec{dx}$ is small compared
to $\vec{x}$.

We define the $\hat{x}$ direction to  be pointing from the
more massive component to its companion, and the $\hat{z}$
direction to be parallel to the angular velocity vector.
We assume that the equilibrium 
configurations have reflection symmetries about the $x-y$ and
$x-z$ planes. We therefore distribute our $N$ points on the 
surface of a quarter-sphere. In the discussion that follows, 
we use $\sim1600$ points on a quarter-sphere for each of the
components.

\section{Equilibrium Figures of Rotation}
\label{results}

\subsection{Physical Review and Analytic Approximations}

In this section, we provide a brief
introduction to the basic physical concepts
underlying the derivation of equilibrium figures of rotations.
We start by introducing the properties of single rotating
Maclaurin spheroids. We then proceed to examine the properties
of infinitesimal spheroids orbiting a massive companion. 
While the exact equilibrium configuration of a satellite
may be far from spheroidal, this exercise captures the basic
properties of the equilibrium figures of binary systems.
We then proceed to examine triaxial ellipsoids.
We write the equations for the single Jacobi ellipsoids,
and then study infinitesimal triaxial ellipsoids satellites
in binary systems.

In section~\ref{numRes} we present our numerical results for the 
general equilibrium configuration of mutually deformed binaries.
Our numerical configurations deviate
from pure ellipsoidal forms.
We discuss the properties of our solutions
and compare them with the previous estimates
based on the Roche ellipsoids in section~\ref{discussion}.

\subsubsection{Maclaurin Spheroids}
\label{MacL}

In his classical derivation of the rotational distortion of
the Earth, Newton argued that since the Earth is in equilibrium,
the weights of liquid filling two canals, one stretching from 
the center of the Earth to the equator, and the other from the
center to the pole, must be equal. This is equivalent to 
demanding that the equator and the pole are a part of the same
equipotential surface. Newton further pointed out that
since the accelerations associated with both gravity
and rotation are proportional to the distance from the center
of the Earth, these weights (or, equivalently, the surface potentials
measured relative to the center) are given by $0.5\;h\;a(h)$, where $h$ is
the height of the surface, and $a(h)$ the acceleration on the 
surface.

When Maclaurin later addressed the general problem of
spheroidal equilibrium configurations, he followed the arguments
outlined by Newton. For the potentials on the equator and at the pole
we have,
\begin{subequations}
\begin{eqnarray}
V_{\rm equator} = \frac{1}{2} a \times \left( g_{\rm equator} - \omega^2 a \right)
\\
V_{\rm pole} = \frac{1}{2} c \times g_{\rm pole}
\end{eqnarray}
\end{subequations}
where $g_{\rm equator}$ is the gravitational acceleration on the equator,
$g_{\rm pole}$ is the gravitational acceleration at the pole, $\omega$
is the angular velocity, $a$ the semi-major axis $(a=b>c)$ of the spheroid,
$c=a\sqrt{1-e^2}$ is the semi-minor axis, and $e$ is the eccentricity.

For a spheroid, the gravitational acceleration along the equator
and at the pole are analytical functions of the eccentricity 
$e$, and are given by 
(e.g. Chandrasekhar~1969),
\begin{subequations}
\footnotesize
\begin{eqnarray}
g_{\rm equator} = 2\pi G\rho a \frac{\sqrt{1-e^2}}{e^3} \left( \sin^{-1}e -e\sqrt{1-e^2} \right)
\\
g_{\rm pole} = 4\pi G\rho a \frac{\sqrt{1-e^2}}{e^3} \left( e - \sqrt{1-e^2}\sin^{-1}e \right).
\end{eqnarray}
\end{subequations}
Requiring that the potentials on the equator and at the pole are equal then yields,
\begin{equation}
\normalsize
\begin{array}{l}
\omega^2 = \frac{g_{\rm equator} - g_{\rm pole} \sqrt{1-e^2}}{a} = 
\\
\pi G\rho \left[\frac{\sqrt{1-e^2}}{e^3}2(3-2e^2)\sin^{-1}e -\frac{6}{e^2}\left( 1-e^2 \right) \right],
\end{array}
\end{equation}
providing the relation between the eccentricity and angular velocity for 
Maclaurin spheroids.

\subsubsection{An Illustrative Example: An infinitesimal Spheroid in
a Binary System}
\label{2spheroids}

Let us now consider the problem of a binary system, comprised of 
a massive primary of mass $M$ orbited by an infinitesimal spheroidal 
($a=b>c$) satellite. Here we assume that the system is tidally locked, so that
the spin and orbital angular velocities are equal.
For an infinitesimal satellite, the system's center of mass is effectively
at the center of the massive primary, and we assume that the satellite 
is in a circular Keplerian orbit.

The massive primary exerts a tidal force on the spheroidal satellite. 
On the point facing the primary, this results in 
an acceleration opposing the satellites' own
gravitational acceleration, of magnitude $2GMa/d^3$, where $d$ is the
separation between the two components.
On the polar axis, the primary serves to compress the secondary,
providing an acceleration along the polar axis of $GMc/d^3$.

Following the previous line of argument, the potentials on the surface
of the secondary can be readily computed.
Writing $V_a$ for the potential on the equatorial point facing the primary, 
and $V_c$ for the polar potential,
\begin{subequations}
\label{spheroidal_v_om}
\begin{eqnarray}
V_a = \frac{1}{2} a \left( g_{\rm equator} - \omega^2 a - \frac{2GMa}{d^3} \right)
\\
V_c = \frac{1}{2} c \left( g_{\rm pole} + \frac{GM}{d^2} \frac{c}{d} \right)
\end{eqnarray}
\end{subequations}
with $c = a \sqrt{1-e^2}$.

Demanding that these two potentials are equal, and assuming Keplerian
rotation, 
\begin{equation}
\label{Kepler}
\frac{GM}{d^3} = \omega^2,
\end{equation}
we now obtain
\begin{equation}
\label{spheroid-eccentricity}
\omega^2 = \frac{g_{\rm equator} - g_{\rm pole}\sqrt{1-e^2}}{(4-e^2)a},
\end{equation}
which relates the eccentricity and rotational velocity of
infinitesimal spheroids in binary systems.
We note that the resulting spheroids are {\it not} genuine equilibrium 
configurations, because even for the eccentricities dictated by
equation~(\ref{spheroid-eccentricity}), for which $V_a=V_c$, the value
of the potential varies along the surface the spheroids.

\begin{figure}[!h]
\epsscale{1.0}
\plotone{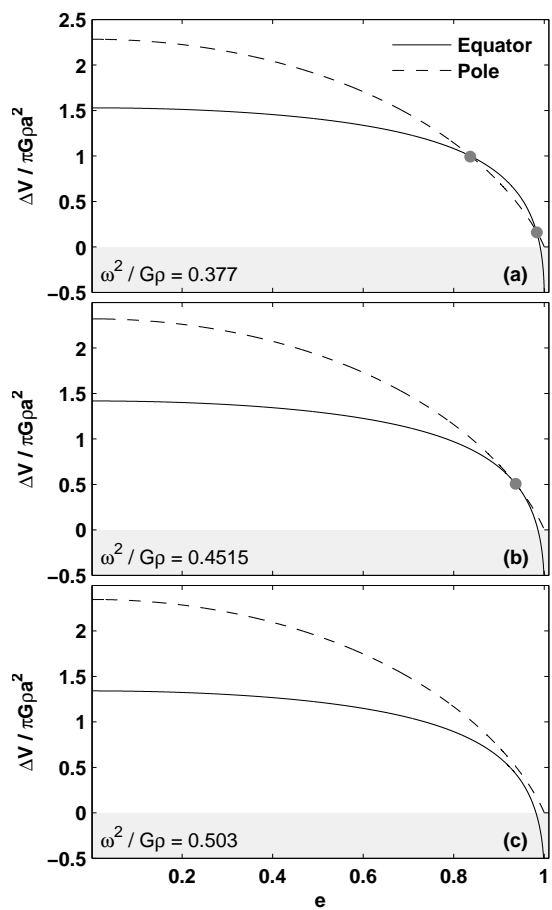}
\caption{The equatorial (solid) and polar (dashed) potentials 
versus eccentricity for an infinitesimal spheroid in a binary system.
An equilibrium solution arises where the two curves cross (filled
circles). The upper panel is for $\omega^2/G\rho = 0.377$, the 
middle is for the ``Roche limit'' spheroid (see text) 
$\omega^2/G\rho = 0.4515$,  and the lower panel is for 
$\omega^2/G\rho = 0.503$. In the shaded area ($\Delta V<0$), the net force
is away from the center.}
\label{compV}
\end{figure}

Figure~\ref{compV} shows the equatorial ($V_a$)
and polar ($V_c$) potentials (relative to the satellite
center, see equation~\ref{spheroidal_v_om}), as functions of
eccentricity for three representative values of the angular
velocity. 
The solid curves show the equatorial potential, and the dashed
curves show the polar potential. Equilibrium configurations arise 
where the two potentials obtain equal values and the curves cross.
This is shown by the filled circles in Figure~\ref{compV}.

For low angular velocities (panel a) there are two eccentricity
solutions for a given angular velocity. The lower eccentricity 
solution is stable. A flattening of the equilibrium
spheroid will result in restoring forces acting to raise the
poles and compress the equator, and vice versa. 
The higher eccentricity solution is unstable. A
flattening of the equilibrium spheroid will result in forces
that act to flatten it even further, driving the eccentricity
away from the equilibrium value.
Figure~\ref{compV} shows that the polar potential is always positive
(relative to the satellite center), and the forces on the
pole are therefore always directed toward the center. 
However, the equatorial potential on the point facing the primary
becomes negative at large eccentricities, resulting in an
outward force acting to unbind the material.

Panel (b) shows that there is a limiting angular velocity
for which only one solution exist. This is known as the
``Roche limit'' configuration. For even higher angular 
velocities (panel c), the potential at the pole is higher
than the potential at the equator for {\it any}
eccentricity, and no equilibrium solution can be found.

Note that at the ``Roche limit'', the potential is still
positive, and the material remains bound. The Roche limit
is therefore crossed before the ``Roche lobe''\footnote{The 
Roche lobe is the region of space around a body within which
a test particle is gravitationally bound to that body.}
is filled.
For larger values of angular velocity, the
force on the pole is {\it always} larger than the force
on the equator (see panel c), and the configuration is
therefore driven to an ever increasing eccentricity.
Eventually, the potential on the equator will become negative,
resulting in mass shedding along the equator.
Even though the Roche limit occurs before the Roche lobe
is filled, once the Roche limit is crossed the configuration 
will evolve to shed mass.

While our example assumed an unrealistic spheroidal configuration
for the infinitesimal satellite, we find that it captures much of the
physics at play in homogeneous binary systems. This includes the 
existence of two solutions, one of which is stable and the other not; 
the existence of a ``Roche limit'' that is reached before the Roche
lobe is filled; and the understanding that despite this fact, beyond
the Roche limit mass shedding will eventually occur. As we discuss 
below, the same principles apply for triaxial configurations.

\subsubsection{Triaxial Ellipsoids: The Jacobi and Roche Solutions}
\label{triaxials}

Triaxial ellipsoids ($a>b>c$) are defined by two eccentricities,
$e_1 = \sqrt{1-(b/a)^2}$ and $e_2 = \sqrt{1-(c/a)^2}$.
To find the equilibrium configurations of single
triaxial rotating ellipsoids, we demand that the surface potential
along the three principal axes be equal. 
This provides two equations, which can be solved for the
two eccentricities.

The gravitational accelerations along the principle axes
of a triaxial ellipsoid are (e.g. Chandrasekhar~1969),
\begin{equation}
g_i = 2\pi G\rho A_i x_i
\end{equation}
where
\begin{equation}
A_i = a_1 a_2 a_3 \int_0^{\infty} \frac{du}{(a_i^2+u)\sqrt{(a_1^2+u)(a_2^2+u)(a_3^2+u)}}\,.
\end{equation}

A comparison of the potential along the semi-major axis $a$,
and the other equatorial axis $b$, yields
\begin{equation}
g_a - \omega^2 a  = g_b\sqrt{1-e_1^2} - \omega^2 a (1-e_1^2)\,,
\end{equation}
whereas for the minor (polar) axis $c$,
\begin{equation}
g_a - \omega^2 a  = g_c \sqrt{1-e_2^2}\,.
\end{equation}
These two equations can be solved numerically
for the values of $e_1$ and $e_2$ given an angular velocity $\omega$. 
These solutions are the Jacobi ellipsoids.

\begin{figure}[!h]
\epsscale{1.0}
\plotone{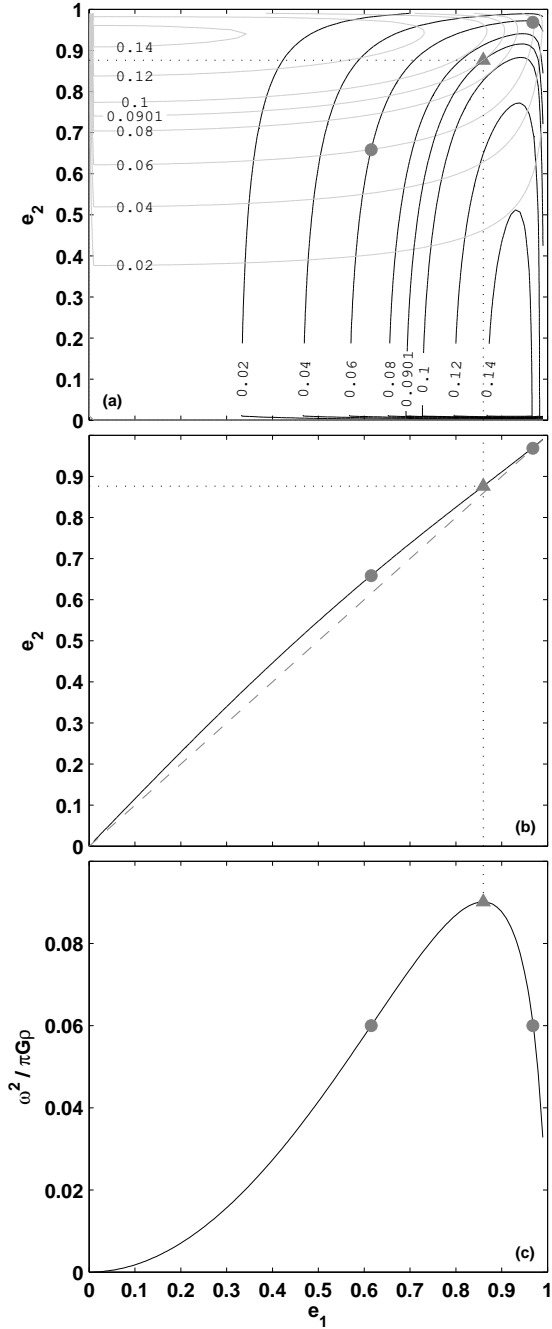}
\caption{Roche Ellipsoid solutions for $q\ll1$. Panel (a) shows the angular
velocities given by equations~\ref{om-roche}, (dark and gray contours) as functions
of the eccentricities $e_1$ and $e_2$. A solution is found when dark and grey
contours of equal values cross. For example, the filled circles show the two
solution obtained for $\omega^2/\pi G\rho=0.06$. The filled triangle is the single
solution obtained for $\omega^2/\pi G\rho=0.0901$. The solid curve in panel (b)
shows the eccentricities of the Roche ellipsoid solutions. The solutions shown 
in panel (a) are displayed here again. Panel (c) shows the angular velocity 
associated with these solutions. The single solution for
$\omega^2/\pi G\rho=0.0901$ has the maximal possible angular 
velocity and is therefore the  ``Roche limit'' ellipsoid.}
\label{RocheE}
\end{figure}

If an infinitesimal triaxial ellipsoid is rotating about a massive spherical
primary, the potential must be modified to include the tidal and gravitational
contributions of the primary.
In this case,
\begin{equation}
\begin{array}{l}
g_a - \omega^2 a  -\frac{2GMa}{d^3} = \\
g_b\sqrt{1-e_1^2} - 
\omega^2 a (1-e_1^2) + \frac{GMa}{d^3} (1-e_1^2)\,,
\end{array}
\end{equation}
and
\begin{equation}
g_a - \omega^2 a -\frac{2GMa}{d^3} = g_c \sqrt{1-e_2^2} 
+ \frac{GMa}{d^3} (1-e_2^2)\,.
\end{equation}

Assuming Keplerian rotation (equation~\ref{Kepler}), 
we may express the above equations
in terms of the eccentricities  $e_1$ and $e_2$, and the 
angular velocity only,
\begin{subequations}
\label{om-roche}
\begin{eqnarray}
\omega^2 = \frac{g_a - g_b \sqrt{1-e_1^2}}{3 a} \\
\omega^2 = \frac{g_a - g_c \sqrt{1-e_2^2}}{(4-e_2^2)a}.
\end{eqnarray}
\end{subequations}
These are the Roche ellipsoidal solutions for an extreme mass ratio ($q\ll1$).
Figure~\ref{RocheE} shows the Roche ellipsoid solutions for $q\ll1$. 
Panel (a) shows the angular velocities given by equations~\ref{om-roche}a and b
(dark and gray contours) as functions of the eccentricities $e_1$ and $e_2$.
Equilibrium solutions exist when the two equations yield equal values
of angular velocity. Examples of equilibrium configurations are shown
by the filled symbols. The circles correspond to
$\omega^2 / \pi G\rho=0.06$, and the triangle to  $\omega^2 / \pi G\rho=0.0901$.

Panel (b) shows the positions of the equilibrium solutions (solid curve).
The dashed curve shows the line $e_1=e_2$ (a prolate spheroid) for comparison.
In panel (c) we display the angular velocity for which an equilibrium solution is found
as a function of $e_1$.
As in the case of the spheroidal satellite, for low angular velocities,
two equilibrium solution are found (see filled circles). 
Panel (c) shows that there exists a maximal value of angular
velocity for which only one solution can be found.
This is the ``Roche limit'' ellipsoid, occurring at
$\omega^2 / \pi G\rho=0.0901$ (for $q\ll1$), shown here by the filled
triangle.
For values of angular velocity beyond the ``Roche limit'', 
no solution can be found for {\it any} combination of eccentricities.

\subsection{Numerical Results}
\label{numRes}

\begin{figure}[!h]
\epsscale{1.0}
\plotone{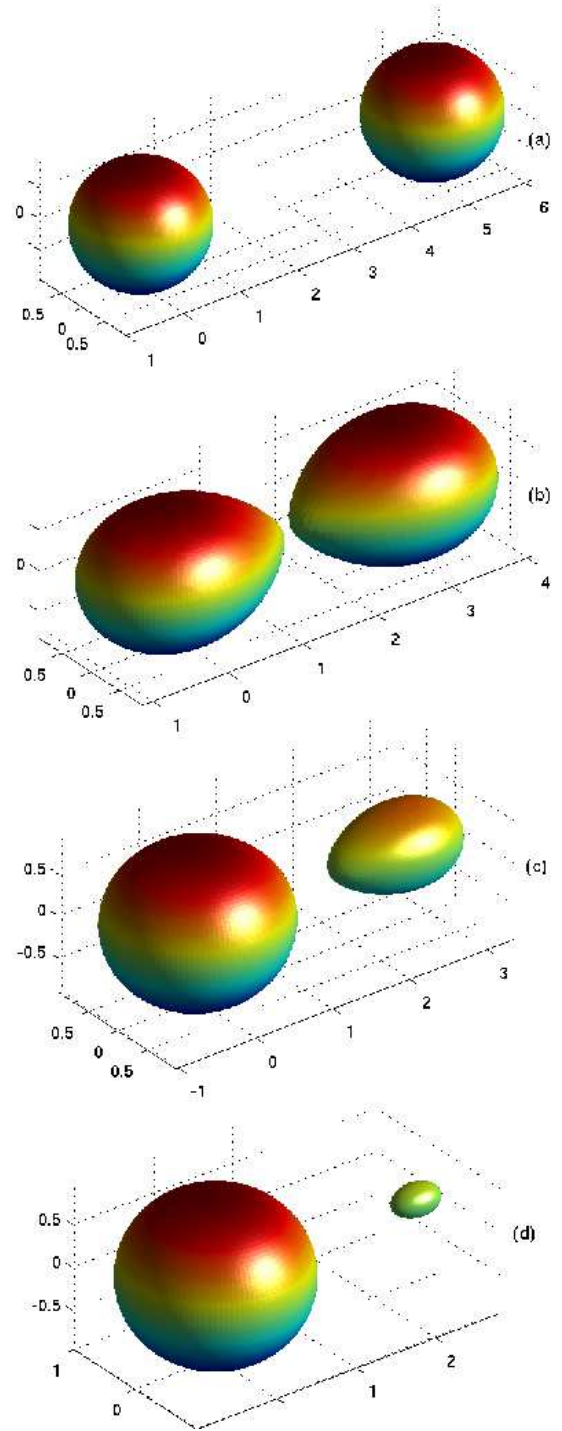}
\caption{Equilibrium configurations of rotating binaries. (a) $q=1$,
distant components. (b) $q=1$, close components. (c) $q=0.2$,
close components. (d) $q=0.01$, close components.}
\label{EGconfs}
\end{figure}

We have computed the equilibrium configurations of rotating binary
systems for mass ratios, $q$, between ~$10^{-3}$~ and ~$1$.
For each value of $q$, we found solutions starting at a very low
angular velocity (large separation), for which the rotational and
tidal deformations are small, and following a sequence of increasing
angular velocity, ending at the Roche limit where an equilibrium
solution can no longer be found.

In this section we present examples of our numerical results for
the equilibrium figures of rotation. 
These results were obtained using the numerical method described in 
section~\ref{method}, which allows the bodies to take {\it any} form that is
symmetric about the $x-y$ and $x-z$ planes\footnote{The bodies are
separated along the $x$-axis, and the rotation is about the $z$-axis.}.
As we demonstrate below, in some cases our numerically computed
equilibrium figures depart from an ellipsoidal form.

In figure~\ref{EGconfs}, we show several examples of equilibrium
configurations.
Panel (a) shows the equilibrium forms of equal-mass components, 
orbiting at a large separation (small angular velocity). The 
equilibrium figures in this case are, as expected, close to spherical.
Panel (b) shows the same system orbiting at a closer separation
(note different scale).
Asphericity becomes apparent for this rapidly rotating system.
The equilibrium configurations resemble triaxial ellipsoids,
but departures from pure ellipsoidal forms can be readily seen.

Panel (c) shows an example of a smaller mass ratio, $q=0.2$.
In this case, the larger body is only mildly deformed,
due to the smaller gravitational perturbation of its low-mass
companion.
However the smaller body is visibly distorted by the gravitational
forces of its high-mass companion. Departures from 
ellipticity are apparent for the lower-mass component.

Finally, in panel (d) we show an extreme mass ratio, $q=0.01$.
For this system the center of mass is close to the center
of the massive component. The lower mass satellite, is significantly
deformed, yet remains nearly ellipsoidal (see Section~4.1). The 
massive component is unaffected by the minute gravity of its low-mass 
companion, and is deformed merely by its own rotation, thus taking 
the form of the oblate Maclaurin spheroid of the appropriate angular
velocity.

\section{Discussion}
\label{discussion}

\subsection{Non-ellipticity}

Consider the gravitational potential that a primary
of mass $M$ creates on the surface of its companion, 
which has a semi-major axis $a$, and is located at
a distance $d$. On the axis connecting the two components,
\begin{equation}
\label{pot-exp}
V(a) = \frac{GM}{d} \left[ 1 - \frac{a}{d} + \left(\frac{a}{d}\right)^2 
- \left(\frac{a}{d}\right)^3 + ...\right].
\end{equation}
The first term is constant, and does not produce any forces. 
The second terms gives rise to the circular orbital motion. 
The third term is the lowest  order tidal force, which produces 
the symmetric deformations that were considered in the analysis
of the Roche equilibrium ellipsoidal configurations (section~\ref{triaxials}).
The fourth term is the first {\it asymmetric} correction
to the tidal potential, and is a factor $a/d$ smaller
than the previous term. Asymmetric deformation are therefore
always small at large separations. In addition, the primary is
not spherical, as it is also deformed by rotation and
by tidal interactions. This affects the gravitational potential,
and deviation from equation~(\ref{pot-exp}) may also contribute to
non-ellipsoidal deformations.

\begin{figure*}
\epsscale{1.0}
\plotone{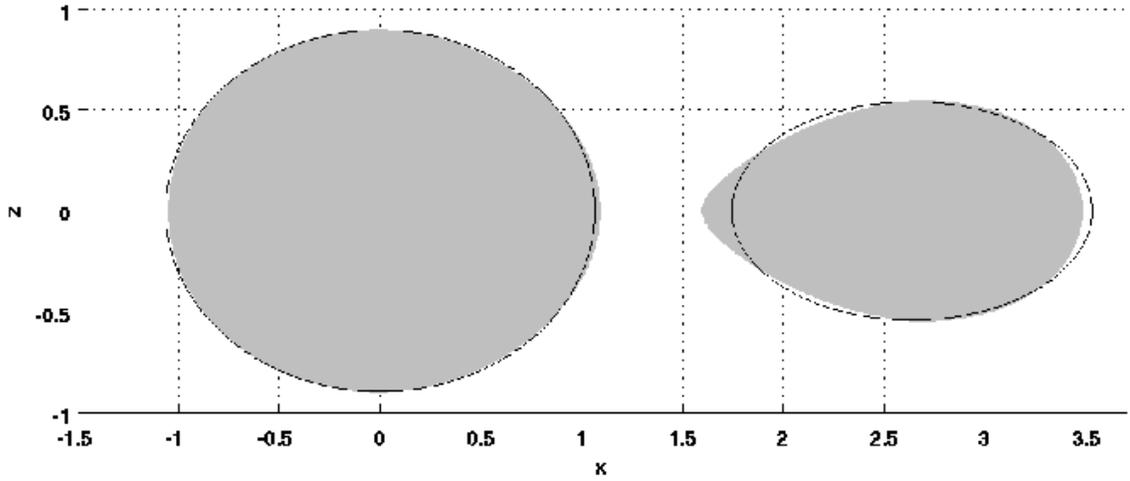}
\caption{Departures from ellipsoidal form, for $q=0.3$, $d=2.636$, 
$\omega^2/G\rho=0.29732$. The shaded area is the projection of
the equilibrium configurations on the $x-z$ plane. The black
contours show a projection of the best-fit ellipsoids.}
\label{EGnone}
\end{figure*}

At a given separation, the semi-major axis of the 
lighter component is $\propto q^{1/3}$. For small
values of $q$, the asymmetric correction to the 
potential is therefore small, and the bodies remain
ellipsoidal to a good approximation. For larger values
of $q$, the asymmetric correction become 
significant\footnote{For eccentric orbits, the average gravitational
potential deviates from a Keplerian potential, not only because of 
the deformations of the body inducing the potential, but also because
of the changing distance along the eccentric orbit. If this distance 
change is much smaller than the deformations, the circular approximation
holds (e.g. for the case of equal mass, near-contact binaries this
implies $e<0.01$). Our models do not apply to higher eccentricities.}.
Figure~\ref{EGnone} shows an example of the departures from
a symmetric ellipsoidal form, for $q=0.3$, $d=2.636$, 
and $\omega^2/G\rho=0.29732$. The shaded area is the projection
of our numerically computed equilibrium configurations on
the $x-z$ plane. The black contours show a projection of
the best-fit ellipsoids. This example shows the asymmetric
deformation of the lighter component. There is a significant
``bump'' on the side facing the primary, while the distant
side is somewhat depressed. The primary's asymmetric deformation
is minor.

\begin{figure*}
\epsscale{1.0}
\plotone{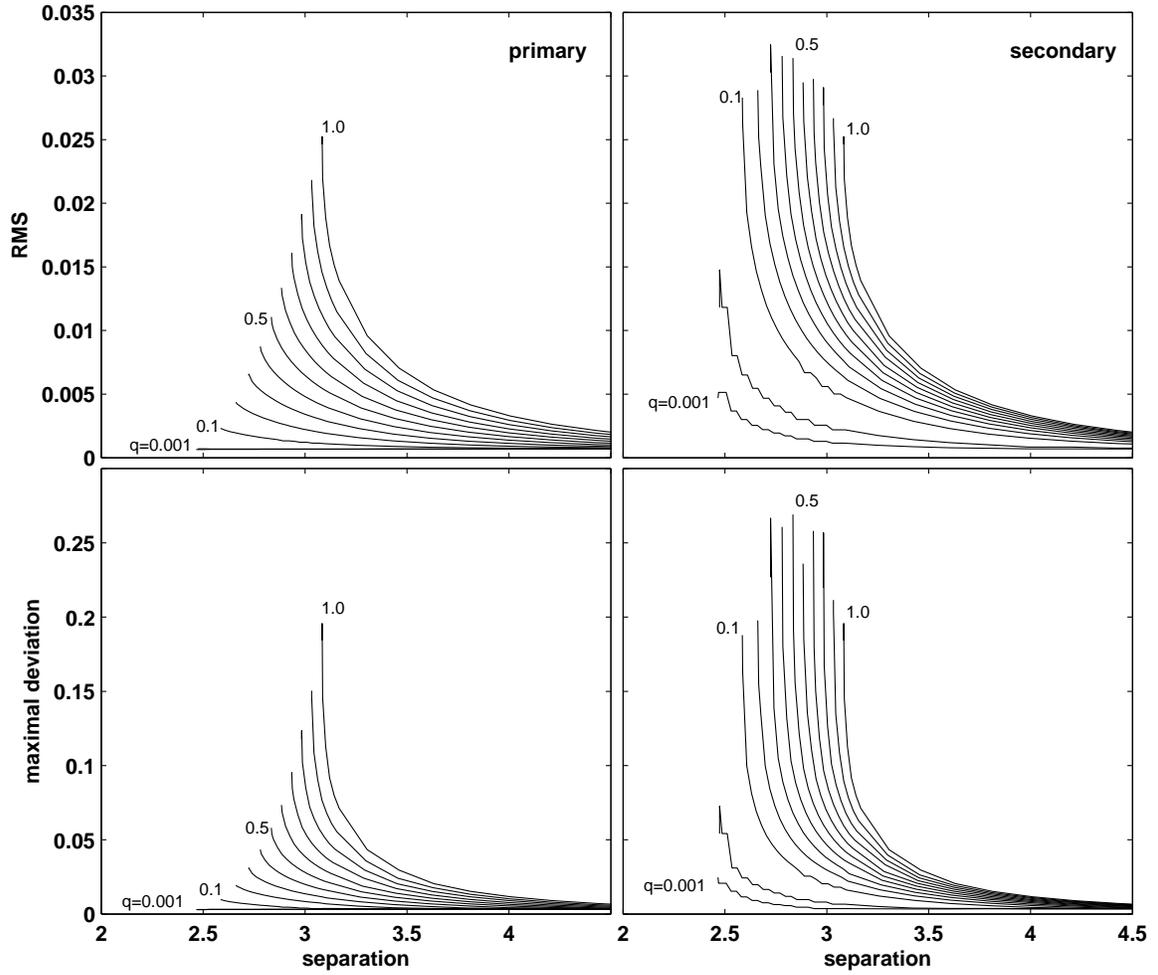}
\caption{Departures from ellipsoidal configurations. 
The upper panels show the RMS of differences in radii
between our numerical configurations and triaxial 
ellipsoids fitted to them, as a function of separation. 
The lower panels show the maximal local deviation 
between the numerical results and the best-fit ellipsoids. 
The left hand panels are for the massive component
(primary), and the right hand panels are for the
secondary. Different curves are for different mass
ratios, $q=10^{-3},\; 10^{-2},\; 0.1,\; 0.2,\; 0.3, ...\; 1$.
Some labels are indicated near the curves for guidance.}
\label{non-e}
\end{figure*}

To estimate the extent to which our equilibrium forms depart
from pure ellipsoids, we fitted an ellipsoid to
each numerically computed configuration, and then
considered the differences between the best-fit ellipsoid
surface and the numerical solution.
In figure~\ref{non-e} we display the departures from ellipsoidal
configurations.
The upper panels show the root mean square (RMS) of differences
between the two forms, $\sqrt{\sum(R_{i,{\rm el}} - R_{i,{\rm sol}})^2/N}$,
where $R_{i,{\rm el}}$ is the distance from the center to the best-fit 
ellipsoid surface, $R_{i,{\rm sol}}$ is the distance from the 
center-of-mass to the surface of the numerical solution, and $i=1,..,N$.
The left hand panel is for the primary, and the right
hand panel for the secondary.
The RMS of the primary (secondary) and the separation were
normalized to the primary's (secondary's) mean radius.
In each panel, different curves are shown for different values of $q$.
The lowest curve is for the smallest mass ratio that we 
consider, $10^{-3}$, followed by  
$q=0.01,\;0.1,\;0.2,...,\;1.0$ in increasing order. 
Some labels are shown near the curves for guidance.

Figure~\ref{non-e} shows that indeed ellipsoids are a good
fit at large separations, and become inaccurate at small
separations. It also shows that for both of the components,
the RMS at a given distance increases as a function of $q$.
At small separations and comparable mass ratios ($q>0.2$)
the typical RMS is $\gtrsim1\%$, and can be as large
as $\sim3\%$ for the most extreme cases.

The departures from ellipsoidal forms are most prominent
in the direction facing the companion (see Figure~\ref{EGnone}). 
The RMS may therefore remain relatively small even in cases
where locally, large deviations occur. In the lower panels 
of figure~\ref{non-e} we therefore show the largest deviation
between the equilibrium configuration and its best ellipsoidal
fit, for the primary (left panel) and secondary (right panel).
At small separations and comparable mass ratios, deviation of
order $10\%$ ($20\%$) occur for the primary (secondary) component.

\subsection{Departures from Keplerian Rotation}

\begin{figure}[!h]
\epsscale{1.0}
\plotone{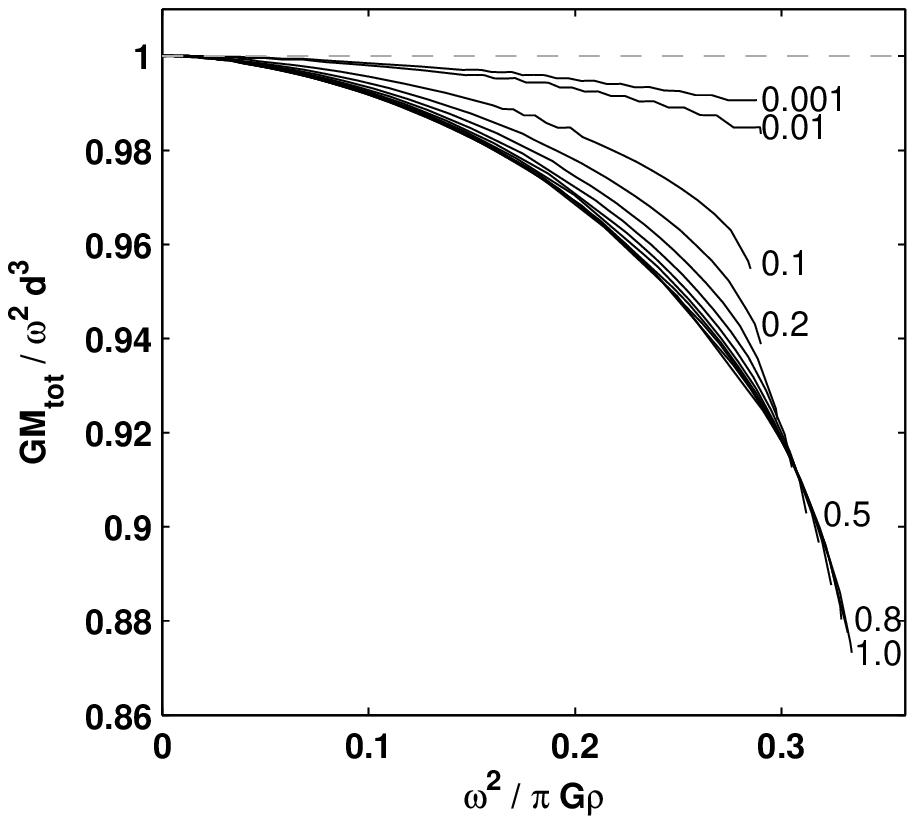}
\caption{Departures (solid curves) from Keplerian rotation (dashed line) 
for mass ratios of $10^{-3}$, $10^{-2}$, $0.1$, $0.2$, ... $1.0$ 
(top to bottom). 
Some labels are indicated near the curves.}
\label{non-kep}
\end{figure}

The tidal and rotational deformations of the equilibrium
configurations modify their gravitational potentials, 
causing them to deviate from those of spherical masses.
These modified potentials affect the dynamics of the orbit,
leading to departures from Keplerian rotation. 
For a given total mass and angular velocity, the separation 
is generally larger than that predicted by Kepler's law.

At large separations (small angular velocities), the bodies
remain approximately spherical and the orbit is nearly
Keplerian.
At small separations (high $\omega$), even for extreme
mass ratio ($q\ll1$), the rotational deformation of the primary
leads to small departures from Keplerian rotation. For larger
values of $q$ the effect grows, as triaxial and higher order
deformations become significant.

Figure~\ref{non-kep} shows the extent to which our 
numerical solutions depart from Keplerian rotation, 
measured in terms of $GM_{\rm tot} / \omega^2 d^3$.
Keplerian rotation is given by the constant line 
$GM_{\rm tot} / \omega^2 d^3 = 1$ (dashed gray line).
The numerical solutions for mass ratios between
$10^{-3}$ and $1$ are shown by the solid curves.
Departures from Keplerian rotation are a growing function 
of the angular velocities for any $q$.
For $q=10^{-3}$, the maximal departure near the 
Roche limit is of order $1\%$.
For comparable masses, departures of order $10\%$ occur
at the Roche limit, reaching a maximal value of $\sim13\%$
for equal mass components.

\subsection{Angular Momentum}

\begin{figure}[!h]
\epsscale{1.0}
\plotone{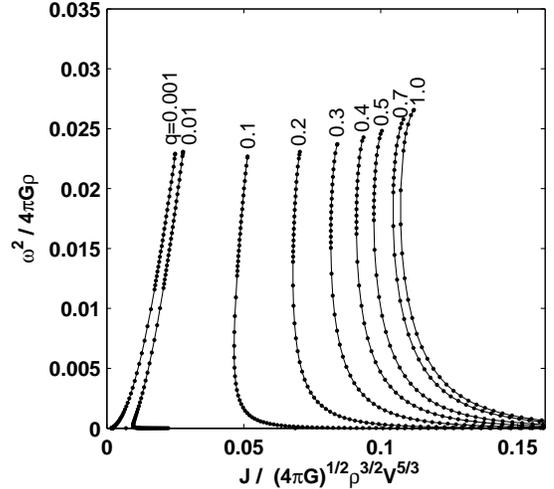}
\caption{Angular velocity versus angular momentum for our numerical 
solutions. Data shown for $q$ between $0.001$ and $1$, as indicated
by the labels. Lines are shown to guide the eye in connecting points
with the same $q$.}
\label{Jqall}
\end{figure}

In figure~\ref{Jqall} we show the angular velocity
($\omega^2 / 4\pi G\rho$) versus the angular momentum
($J / \sqrt{4\pi G} \rho^{3/2} V^{5/3}$, where $V$ is the total
volume). Different curves are
for different mass ratios, between $10^{-3}$ (leftmost curve)
and $1$ (rightmost curve).

At a given mass ratio, the angular momentum first 
decreases as $\omega$ grows, but later begins to
increase again. Each curve ends at a ``Roche limit''
appropriate for its mass ratio,  where an equilibrium
configuration can no longer be found.

\begin{figure}[!h]
\epsscale{1.0}
\plotone{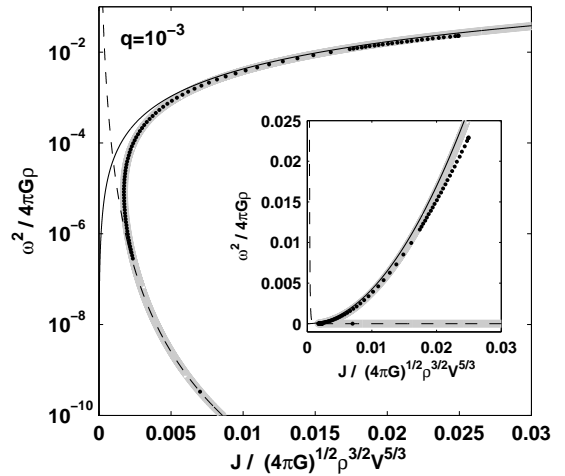}
\caption{Angular velocity versus angular momentum.
The solid and dashed lines show the analytic expressions for the 
spin angular momentum of the primary, and orbit angular momentum 
of the secondary for $q\ll1$. The thick gray line is the sum of
both components. The points show the results of our numerical
computations for $q=0.001$. The inset shows the same data in 
linear scale.}
\label{Jq0.001}
\end{figure}

Focus first on the data for $q=10^{-3}$, shown again in 
Figure~\ref{Jq0.001}. For this extreme mass ratio, the
system is rotating about the primary's center.
Consider a primary of mass $M$ and radius $R$, orbited 
by a satellite of mass $m=qM$ at a distance $d$.
The angular momentum is the sum of two terms,
an angular momentum due to primary spin,
\begin{equation}
\label{Js}
J_{\rm spin} \simeq \frac{2}{5} M R^2 \omega
\end{equation}
and an angular momentum due to the secondary orbit,
\begin{equation}
\label{Jo}
J_{\rm orbit} \simeq m d^2 \omega.
\end{equation}
For Keplerian rotation, $J_{\rm orbit} \propto \omega^{-1/3}$.

At low angular velocities, the separation between the
two components is large, and the orbital angular momentum
dominates. As the angular velocity increases, $J_{\rm orbit}$ 
decreases, until the contribution of the primary's
spin angular momentum begins to dominate, and the 
angular momentum is then proportional to $\omega$.

This is shown in figure~\ref{Jq0.001}.
The dashed line shows the orbital angular momentum
(equation~\ref{Jo}) that dominates at small angular velocities.
The dark solid curve is the primary spin angular momentum
(equation~\ref{Js}) that dominates at large $\omega$.
The thick gray lines is the sum of both contributions.
This simple analytic approximation nicely reproduces the
trend followed by the numerical results.

For less extreme mass ratios, the spin and orbit contributions
due to both components must be taken into account.
A simple analytical approximation may be derived under 
the simplifying assumptions of Keplerian rotation,
and a known deformation.
Here we assume that the axes-ratios for both components
of the binary system, and for both axes in each component
are equal, such that
$b_1/a_1 = c_1/a_1 = b_2/a_2 = c_2/a_2=\sqrt{1-e^2}$,
where $a_i>b_i>c_i$ are the principle axes of the $i$'th
components.
In this approximation,
\begin{equation}
J_{\rm spin} = \frac{1}{5} M_1 a_1^2 (1+q^{5/3})(2-e^2) \omega
\end{equation}
and
\begin{equation}
J_{\rm orbit} = G^{2/3} M_1^{5/3} \frac{q}{(1+q)^{1/3}}\omega^{-1/3}.
\end{equation}
The total angular momentum is then given by
\begin{equation}
\label{Jtot}
J = J_{\rm spin} + J_{\rm orbit}.
\end{equation}

At low angular velocities (large separations) the orbital angular
momentum dominates, and $J$ decreases with increasing $\omega$. 
However, as the separation decreases, the contribution of the spin
angular momentum increases, and $J$ grows again.

\begin{figure}[!h]
\epsscale{1.0}
\plotone{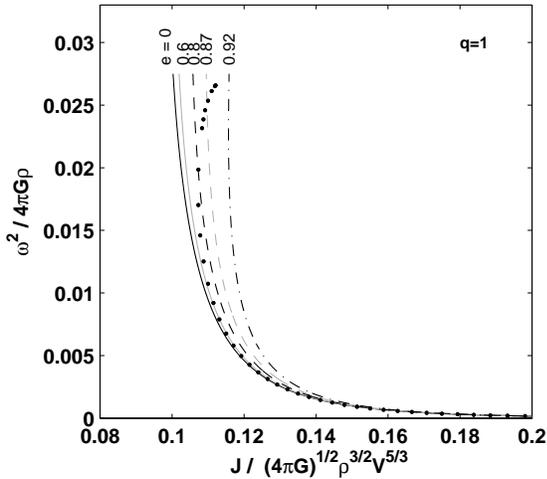}
\caption{Angular velocity versus angular momentum for $q=1.0$.
The points show our numerical results. The different curves show
the analytic approximation of equation~\ref{Jtot} for various values of
eccentricity ($e=0,0.6,0.8,0.87,0.92$), as indicated by the labels.
As $\omega$ increases tidal deformations grow, and the value of $J$ 
continuously shifts to curves of higher eccentricity.}
\label{Jq1}
\end{figure}

For extreme mass ratios ($q\ll1$), the larger component
dominates the spin angular momentum, and the smaller
component dominates the orbit angular momentum.
The asphericity is limited to small (spheroidal)
rotational deformations, and $e$ remains
close to $0$. Equation~\ref{Jtot} thus agrees
with its simplified version above (equations~\ref{Js}
and \ref{Jo}). As the value of $q$ grows, the tidal 
forces deform the bodies, and the configurations 
grow increasingly aspherical.

In figure~\ref{Jq1} we focus on the angular momentum
for $q=1$. The data points show our numerical solutions, and 
the different curves show the analytical approximations of
equation~\ref{Jtot} for $q=1$ and for increasing values
of eccentricity.
The leftmost solid curve is for spheres ($e=0$),
and curves to the right are for higher eccentricities.
At small angular velocities (large separations), the tidal
and rotational deformations remain minor, and the spherical
approximation agrees with the numerical results. 
As $\omega$ increases, tidal forces deform the bodies
into increasingly elongated shapes. 
Figure~\ref{Jq1} indeed shows that as $\omega$ increases, the
values of $J$ grow further and further away from the spherical
$e=0$ curve. In fact, as $\omega$ increases the values of $J$
continuously shifts to curves of higher and higher eccentricity.

Returning now to figure~\ref{Jqall}.
The angular velocity at which the angular momentum achieves its
minimum is an increasing function of $q$.
According to equation~\ref{Jtot}, this occurs at 
$\omega^2 \propto q^{3/2}(1+q)^{-2}(1+q^{5/3})^{-3/2}$.
The angular momentum of equal mass binaries therefore
``turns around'' at a higher value of $\omega$ than for binaries
with extreme mass ratios.

The approximation of equation~\ref{Jtot} thus
explains the existence of a minimum to the angular
momentum, the position of this minimum as a function of $q$,
and the ``drift'' of the angular momentum to values
larger than those expected for spheres.

\section{Light Curves}
\label{secLC}

We have carried out computations of the light curves of Kuiper Belt
binaries, based on the equilibrium figures of rotation presented
in Section~\ref{numRes}. 
The light curves depend on the relative positions of the observed binary, 
the observer, and the light source; on the inclination of the orbit relative
to the plane of the sky; and on the reflecting properties of the materials
composing the surface of the observed system.

For binaries in the Kuiper Belt the observed geometry is simple. At a 
distance of $\gtrsim40$~AU, the observer (on Earth), light source (the Sun), 
and the observed Kuiper Belt binary are almost aligned, so that the 
``phase-angle'' between the line-of-sight and the KBO-Sun direction 
always remains small ($<2^\circ$).

The reflecting properties of  KBOs are not well constrained.
Here we consider two simple options.
First, we consider uniform reflection, which produces an observed intensity
proportional to the projected area on the plane of the sky. For the given geometry,
this can result from simple ``backscatter'' reflection\footnote{The Lommel-Seeliger
law also reduces to being proportional to the geometrical cross-section at low
phase angles (e.g. Lacerda \& Jewitt~2007).}.
Second, we study the case of diffuse (``Lambertian'') reflection, for
which light is reflected equally in all directions, and the observed
intensity from a surface area depends only on the cosine of the angle
between the Sun and the normal to the reflecting surface.

The absolute intensity of light as seen from Earth also depends on the distances
between the Sun, the KBO, and Earth, and on the KBO's size and albedo. 
Here we ignore the absolute intensities and instead focus on the normalized
relative light-curves.
We tile the surface of a given equilibrium configuration with surface triangles,
and then compute the light intensity for a given orbital phase, inclination, and
reflection law, taking into account possible obscurations of different 
tiles. The numerical procedure is described in detail in~\ref{LCequations}.
We have verified that our results reproduce previous light-curve computations
given similar configurations (see Appendix~\ref{sanity}).

\begin{figure}[!h]
\epsscale{1.0}
\plotone{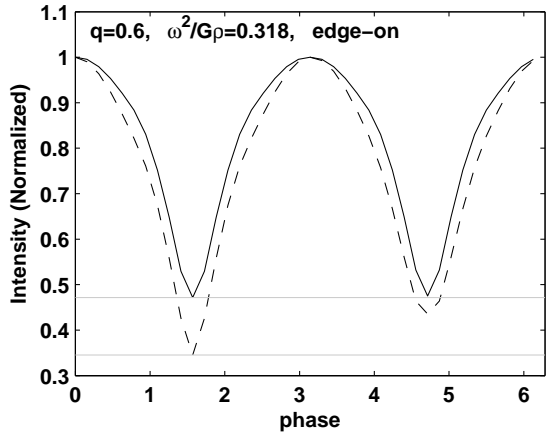}
\caption{Light curves of an edge-on binary system with $q=0.6$ and 
$\omega^2/G\rho=0.318$. The solid curve is for backscatter reflection,
and the dashed curve is for diffuse reflection.}
\label{LCex}
\end{figure}

In Figure~\ref{LCex} we show an example of a light curve computed
for a mass ratio $q=0.6$, an angular velocity $\omega^2/G\rho=0.318$, and
an edge-on orbit. The solid curve shows the light curve for backscatter
reflection, and the dashed curve is for diffuse (Lambertian) reflection.
Both light curves have been normalized so that the intensity equals $1$
at the maximum (note that for a given albedo the absolute intensity is
larger for backscatter reflection).

The light curve shape depends on the orbital parameters. 
In large separation (small $\omega$) binaries, the eclipses only
span a small fraction of the orbit, and the light curves generally appear
to be slowly varying outside of the eclipses, as the projected area
gradually changes. In close binaries (large $\omega$), such ``plateaus'' do
not exist between the two eclipses.
For the mass ratio considered in Figure~\ref{LCex}, $\omega^2/G\rho=0.318$ 
corresponds to a close binary, and the eclipses indeed appear to span most 
of the orbit.

The mass ratio determines the depth of the eclipses. Even for the case
of backscatter reflection, for which the intensity is proportional
to the projected area, eclipses are deepest for equal mass components.
In this case the tidal deformations are largest, and the projected area
is a strong function of orbital phase. The deformations for extreme-mass ratio
binaries are smaller (and more spheroidal), and the relative depth of the
eclipses is thus smaller. For diffuse reflection the impact of the mass 
ratio is even larger. One of the most indicative differences between backscatter
and diffuse reflections, is the relative depth of the two minima.
In backscatter reflection, which is sensitive only to the total projected
area, the depths of the two minima are always equal, regardless of
whether the small body is in front of the large body or vice versa.
However for diffuse reflection, which is sensitive to the surface
curvature, the two minima exhibit different depths. When the small
body in viewed in front of the large body, a larger fraction of the 
surface is inclined relative to the light source, and so the reflected
intensity is smaller.
An additional parameter that affects the depth of the eclipses is the
orbital inclination. Figure~\ref{LCex} shows the maximal variation that
obtains for an edge-on orbit. Higher inclinations result in smaller
variations.

\begin{figure}[!h]
\epsscale{1.0}
\plotone{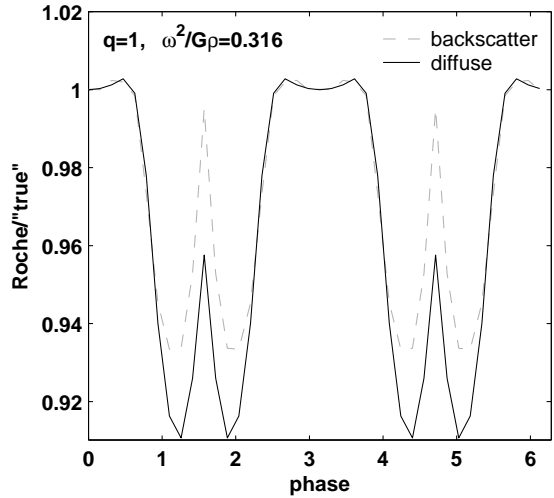}
\caption{The ratio between the Roche-approximation and numerically
computed-configurations light intensities as a function of orbital
phase, for an equal mass binary with $\omega^2/G\rho=0.316$.
In both cases we assumed an edge-on orbit. The solid curve is for diffuse
reflection, and the dashed curve is for backscatter. The maxima of the 
individual light-curves occur at a phase of $0$ and $\pi$.}
\label{compRoche}
\end{figure}

Next we examine the differences between light curves computed assuming
the Roche approximation and light curves computed using our 
numerical solution. As an example, we consider the light curve of an
equal mass binary, with an angular velocity $\omega^2/G\rho=0.316$.
In Figure~\ref{compRoche} we show the ratio between the light intensity
resulting from the Roche configuration appropriate to this angular 
velocity, and our non-ellipsoidal non-Keplerian configurations. 
Both light curves were computed using the numerical
procedure described in Appendix~B, assuming edge-on inclination
and diffuse reflection. They differ only by the equilibrium 
forms. The phase is defined so that the maxima of each individual light 
curve is obtained at $0$ and $\pi$.

Figure~\ref{compRoche} shows that light curves differ by up to $\sim10\%$,
and that the ratio between the two cases is a function of orbital phase. 
Intensity variations of $\sim10\%$ can be easily detected with current
facilities, and the numerical configurations may thus produce superior
fits to observed data, and provide better constrains on the physical
parameters of the observed system.

\section{Application to Kuiper-Belt Binary 2001~QG$_{\bf 298}$}
\label{QGfit}

\subsection{Observational Review}
2001~QG$_{298}$ has been discovered as a part of the Hawaii Kuiper
Belt Variability Project (Jewitt \& Sheppard~2002; Sheppard \&
Jewitt~2002, 2003). The observations, carried out with the 
University of Hawaii $2.2$~m telescope and with the Keck $10$~m
telescope, are reported in Sheppard \& Jewitt~(2004;
hereafter SJ04).
SJ04 indicate that 2001~QG$_{298}$ has a peak-to-peak
photometric range of $1.14\pm0.01$~mag in the {\it R}-band, and a 
period of  $6.8872\pm0.0002$~hours for a single-maximum period
(which may arise due to albedo variations), or
$13.7744\pm0.0004$~hr for a double-maximum period (which may
arise due to rotation).
SJ04 note that the double-peaked light curve provides
a better fit to the observed data, for which the two minima differ
by about $0.1$~mag.
Furthermore, Keck {\it VBR} colors of 2001~QG$_{298}$ indicate 
no color variations along the period, within the photometric 
uncertainties of a few percent.

The observed amplitude variation of $1.14$~mag is exceptionally large
among large ($>25$~km) solar-system objects.
SJ04 discuss three possible reasons for the
large photometric range. First, they consider the possibility of albedo 
variations, and conclude that this scenario is unlikely. On asteroids,
albedo brightness variations are usually smaller than $10-20\%$. Larger albedo
variations are likely to be associated with color variations
(e.g. Iapetus).
Finally, the fact that a double-peaked period provides a better fit
with two distinct minima favors a light curve produced by rotation
rather than by albedo variations.

Next, they consider the possibility of an elongated shape. SJ04 use the
observed photometric range to infer an axis ratio $b/a=0.35$ (assuming the
intensity is proportional to the projected area). Given the absolute 
luminosity of 2001~QG$_{298}$, they derive a semi-major axis between $170$ 
and $270$~km for ``typical'' KBO albedos range ($0.04-0.1$). 
Bodies of this size are
unlikely to be held by material strength over long time-scales. 
Gravity-dominated bodies are rotationally deformed into Maclaurin 
spheroids or Jacobi ellipsoids, depending on their angular velocity. 
SJ04 note that the maximal photometric range for
stable Jacobi ellipsoids is $0.9$~mag, lower than the observed
variation of 2001~QG$_{298}$. This maximum variation occurs for a Jacobi ellipsoid
with $b/a=0.432$ and $c/a=0.345$ (Chandrasekhar~1969; Farinella et al.~1981)
and more elongated objects are dynamically unstable. 
However, we note here that the maximum variation quoted in SJ04 applies
to backscatter reflection, whereas for diffuse (Lambert) reflection 
these axis-ratios yield a variation of $1.5$~mag.
SJ04 also note that the rotational period of 2001~QG$_{298}$ is too
long to cause significant elongation for reasonable densities. They conclude
that 2001~QG$_{298}$ is unlikely to be a single rotating object.

Finally, SJ04 consider the possibility of a close 
binary configuration. As discussed above, the components are distorted
both by rotation and tidal forces. The photometric range of 
2001~QG$_{298}$ is consistent with that of a comparable mass close
binary Roche configuration (Leone et al.~1984). 
They conclude that given the large
amplitude variation, long period, and difference between the two minima, 
a close binary is the most likely explanation for 2001~QG$_{298}$.

With the binary scenario, several attempts have been made to derive
the physical properties of 2001~QG$_{298}$ using the Roche ellipsoidal
approximations. SJ04 used the results presented in
Leone et al.~(1984) to estimate the density, and find $\rho\sim1$~g~cm$^{-3}$.
Takahashi \& Ip~(2004) then constructed specific Roche binary light-curve
simulations to fit the observed light curves. Their best fit Roche solution
implies that 2001~QG$_{298}$ consists of two components with a mass
ratio of $0.65$. Their primary has axis-ratios $b/a=0.79$ and $c/a=0.62$, 
and their secondary has $b/a=0.61$ and $c/a=0.56$. The separation is $2.1$
times the primary's semi-major axis (see illustration in Appendix~\ref{sanity}).
This solution indicates a ``mixed'' reflection pattern, with $\sim70\%$ 
diffuse reflection and $\sim30\%$ backscatter (uniform) reflection. For
these parameters, they infer a bulk density of $0.63\pm0.20$~g~cm$^{-3}$.

Later, Lacerda \& Jewitt~(2007) presented an additional model for
2001~QG$_{298}$. In their solution the mass ratio is $0.84$. 
Their primary has $b/a=0.62$ and $c/a=0.65$, and their
secondary $b/a=0.45$ and $c/a=0.41$. The separation is $2.1$ times 
the primary semi-major axis (see illustration in Appendix~\ref{sanity}).  
Lacerda \& Jewitt find that backscatter reflection best fits
the observed data, and infer a bulk density of 
$0.590^{+0.143}_{-0.47}$~g~cm$^{-3}$.
Lacerda \& Jewitt also considered the possibility that 
2001~QG$_{298}$ is in fact a single Jacobi ellipsoid with 
diffuse (Lambert) or backscatter reflection. They conclude that 
the Roche models fit the data significantly better than Jacobi 
ellipsoids.

In the next section, we apply our numerical models to
2001~QG$_{298}$, and demonstrate how our more accurate solutions
can be used to constrain the physical properties of KBO binaries.

\subsection{Light-curve Fitting and Analysis}

We are interested in finding a physical configuration consistent
with the observations of 2001~QG$_{298}$. To this end, we compare the
light curves associated with the equilibrium configurations computed in 
Section~\ref{results} with the observed light curve of 2001~QG$_{298}$ 
(SJ04), after correcting for the light travel-time
and phase angle effects.

To find the best fit, we create a library of light curves for comparison
with observations. However, because the light curve calculations 
are computationally expensive,  we approached the fitting procedure
in two steps. First, we created a library of ``low-resolution 
configurations''. Here we used equilibrium configurations computed
with $200$ patches on a quarter-sphere, to construct light curves
spanning the entire parameter space studied in Section~\ref{results}.
Given this library of light curves, we found the best fit
(in the sense of minimum $\chi^2$) model. 

For each model, which corresponds to a specific combination of 
mass ratio, rotational velocity, and inclination, we fitted two
parameters: $\alpha$ and $\beta$. These are overall normalization
factors for a ``mixed'' light curve composed of both backscatter
and diffuse reflection, such that the total intensity is~
$I = \alpha\,I^{\rm Backscatter}_{\rm model} +\beta I^{\rm Diffuse}_{\rm model}$.
~In addition to these two normalization parameters, we allowed
for a relative constant phase-shift between the observed and
computed light curves. 

\begin{figure}[!h]
\epsscale{1.0}
\plotone{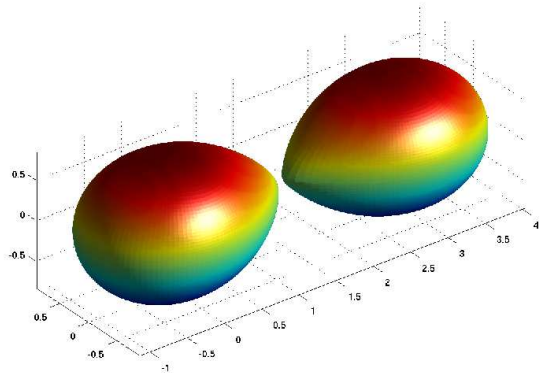}
\caption{The best-fit configuration for 2001~QG$_{298}$.}
\label{bestfit}
\epsscale{1.0}
\end{figure}

Given the best fit solution for the ``low-resolution configurations''
library, we computed a better-sampled, dense library of light curves
based on ``high-resolution configurations'', which have $1600$ patches
on a quarter sphere. For the high resolution configurations, we focus
on the parameter-space surrounding the ``low-resolution''  solution. 
We again searched for the best-fit solution among this new set of model
light curves.

The best fit solution is illustrated in Figure~\ref{bestfit}.
This model has a mass ratio, $q=0.93$, and a rotational velocity 
$\Omega^2/G\rho=0.333$. The best fit is found for pure diffuse
(Lambertian) reflection (so that $\alpha=0$), viewed at an inclination
of $3^o$. For the observed period of 2001~QG$_{298}$, these parameters
imply a density $\rho=0.72$~g~cm$^{-3}$.
The observed magnitude then implies masses of $1.70\times10^{18} (A/0.1)^{-1.5}$
and $1.58\times10^{18} (A/0.1)^{-1.5}$~kg for the two components, where $A$ is
the albedo. While our bodies are not ellipsoidal, an ellipsoidal fit yields
semi-major axes of $102\times78\times71$ and $102\times75\times69$~km$^3$.

\begin{figure}[!h]
\epsscale{1.0}
\plotone{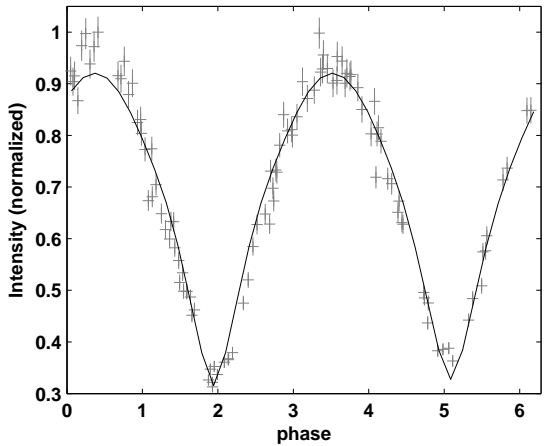}
\caption{Observed light curve for QG$_{\rm 298}$
(data points) and best fit solution (solid curve), 
obtained for
$q=0.93$, $\Omega^2/G\rho=0.333$, with Lambertian 
reflection at an inclination of $3^o$.}
\label{QG}
\end{figure}

We display the observed light curve (symbols) along with the best fit
model (solid curve) in Figure~\ref{QG}. For our assumed intensity errors
of $3.5\%$ on the computed light curves and $4\%$ on the observed data,
this models yields a $\chi^2$ of $123.4$ for $107$ degrees of 
freedom\footnote{For $112$ observed data points and $5$ fitted 
parameters, namely the mass ratio, angular velocity, inclination, values 
of $\alpha$ and $\beta$, and the global phase shift.}.

To estimate the errors on the best fit parameters, we inspect the
values of $\chi^2$ obtained for other models. With five fitted parameters
and a minimal $\chi^2$ of $123.4$, models with $\chi^2<129.3$ are within the
$1\sigma$ ($68.27\%$) error region. Inspecting the model-library, we derive 
the following $1\sigma$ error regions: For the mass ratio, $0.9<q<1.0$; For 
the angular velocity, $0.332<\omega^2/G\rho<0.334$; And for the inclination, 
$2<i<4^\circ$. We further find that all acceptable models have ``pure'' 
diffuse reflection. Given the observed period (and error), the error on the 
angular velocity can be used to compute the error on the bulk density. We 
find a $5.6\%$ error on the density, $\rho=0.72\pm0.04$~g~cm$^{-3}$.

\section{Summary}

In this paper we introduce a numerical method for computing the
self-consistent equilibrium configurations of tidally-locked
homogeneous binaries, rotating in circular orbits. The equilibrium
configurations depend on the mass ratio, angular velocity, and
separation between the two components.

We explicitly compute the gravitational and rotational potentials
on the surface of the two components, and use a Newton-Raphson-based
scheme to converge to an equilibrium solution, for which the
surfaces of each of the bodies is an equipotential surface. 
Our numerical procedure is described in detail in 
Section~\ref{method} and~\ref{equations}.

In Section~\ref{results} we discuss the properties of the equilibrium
figures of rotation. We begin with simple analytic approximations that
we use to study the characteristics of the equilibrium solutions.
We confirm that for low angular velocities (large
separations), two equilibrium configurations always exist, differing
by their eccentricity. The low eccentricity solution is stable. A 
flattening of the low-eccentricity configuration results in restoring
forces. The high-eccentricity solution is unstable. A flattening of
the equilibrium configuration results in forces that act to flatten
it even further, driving the eccentricities away from the equilibrium 
values.

As the angular velocity increases, there exists a ``limiting'' angular
velocity for which only one equilibrium solution exists. This is known
as the ``Roche limit'' configuration. For even higher angular velocities,
no equilibrium solution can be found. At the Roche limit, the material
is still bound, and so the Roche limit is crossed before the Roche lobe
is filled. However, once the Roche limit is crossed, the configuration 
is driven to an ever increasing eccentricity, and eventually mass-shedding
will occur as material flows beyond the Roche-lobe.

We find numerical solutions for the equilibrium configurations
for mass ratios, $q$, between $10^{-3}$ and $1$. For each value of
$q$ we find solutions starting at a very low angular velocity (large 
separation), for which the rotational and tidal deformations are small,
and following a sequence of increasing angular velocity, terminating at the
Roche limit where an equilibrium configuration can no longer be found.

Our numerical solutions indicate that the equilibrium configurations
are not always ellipsoidal. Ellipsoidal fits are generally good approximations
at large separations and for extreme mass ratios. Even at small separations
and for comparable masses, the typical RMS deviation between a 
numerical solution and an ellipsoid fitted to it is $1-3\%$.
However, departures from ellipsoidal forms are most pronounced in the 
direction facing the companion, and the maximal local deviations from 
ellipsoidal forms are of order $10\%$  ($20\%$) for the primary
(secondary) component.

The tidal and rotational deformations of the equilibrium configurations
modify their gravitational potentials. This, in turn, affects the dynamics
of the orbit, leading to departures from Keplerian rotation. 
We measure departures from Keplerian rotation in terms of 
$GM_{\rm tot}/\omega^2d^3$.
At large separations the bodies remain approximately spherical,  
departures from Keplerian rotation remain small, and
$GM_{\rm tot}/\omega^2d^3\sim1$ as it should. At small separations, departures
from Keplerian rotation appear. For $q=10^{-3}$ the maximal deviation near
the Roche limit is $\sim1\%$. For comparable masses ($q>0.5$),
deviations of order $10\%$ occur at the Roche limit, reaching a maximal
value of $\sim13\%$ for equal mass components.

We inspect the angular momenta of the equilibrium configurations.
We decompose the angular momenta into ``spin'' and ``orbit'' components,
and explain the trends in the angular momentum-angular velocity 
parameter-space. As the angular velocity increases, it shifts from being 
orbit-dominated to being spin-dominated, creating a minimum in the 
angular momentum. We discuss the increasing value of this minimum with
increasing mass ratio, and demonstrate the ``drift'' of the angular
momentum to higher values as the tidal deformations grow.

In section~\ref{secLC}, we calculate the light curves that arise
from our numerically computed equilibrium configurations, if placed
in the Kuiper Belt. In addition to the configurations, the observed 
light curves also depend on the inclination of the orbit relative to
the plane of the sky, and on the reflecting properties of the
surface of the objects.

We consider two possibilities for the reflecting properties.
First we consider an observed intensity that is proportional to
the projected area on the plane of the sky, as is appropriate
for backscatter reflection from Kuiper Belt objects.
Second, we consider the possibility of diffuse (Lambert) reflection,
in which the reflected light from a surface area is proportional to
the cosine of the angle between the sun and the normal to the surface.
We note, that backscatter reflection is sensitive only to the total
projected area, and therefore always yields two equal minima. However,
diffuse reflection, which is sensitive to the surface curvature,
generally produces two minima of different depths. When the small body 
is viewed ``in front'' of the large body, a larger fraction of the
surface is inclined.

We compare our light curves to those resulting from the classical
Roche ellipsoidal approximations, and find phase-dependent
intensity deviations of $\lesssim10\%$ between the two cases.

Finally, in Section~\ref{QGfit} we apply our numerical models to Kuiper
Belt binary 2001~QG$_{298}$. This object exhibits an extremely large 
photometric range of $1.14\pm0.01$~mag ({\it R}-band), and a double-peaked
period of $13.7744\pm0.0004$~hr. It is believed to be an example
of a close-binary Kuiper Belt population (SJ04).

Our numerical models indicate that 2001~QG$_{298}$ is well
fitted by a binary with a mass ratio $q=0.93^{+0.07}_{-0.03}$,
an angular velocity $\omega^2/G\rho=0.333\pm0.001$, a nearly
edge-on inclination, $i=3\pm1^\circ$, and pure diffuse reflection. 
For the observed period of 2001~QG$_{298}$, these parameters imply
a bulk density, $\rho=0.72\pm0.04$~g~cm$^{-3}$

SJ04 estimate that 2001~QG$_{298}$ is a representative of a large
Kuiper Belt population of nearly-contact binaries, and 
that at least $10-20\%$ of all large KBOs are, in fact, close-binaries. 
Upcoming LSST observations will identify $>20,000$ new KBOs (LSST Science 
Collaborations:~Abell et al.~2009).
If indeed a significant fraction of the large KBO population is in the form
of contact binaries, the models and methods outlined in this paper may become
essential for the interpretation of their light curves.

\section*{Acknowledgments}

We thank Oded Aharonson, Peter Goldreich, Ehud Nakar, Eran Ofek, David
Polishook, and Hilke Schlichting for helpful discussions. Some of the 
numerical calculations presented in this work were performed on Caltech's 
Division of Geological and Planetary Sciences Dell cluster. 
We thanks Oded Aharonson for his assistance with our use of the cluster.
O.G. acknowledges support provided by NASA through Chandra Postdoctoral
Fellowship grant number PF8-90053 awarded by the Chandra X-ray Center, 
which is operated by the Smithsonian Astrophysical Observatory for NASA 
under contract NAS8-03060.
R.S. research is supported by IRG and ERC grants, and a Packard
Fellowship.

\appendix
\section{Basic Equations and Numerical Procedure}
\label{equations}

\subsection{Definitions}
\label{A1}
We are interested in finding self consistent
equilibrium configurations of a tidally locked binary systems,
comprising of two homogeneous components, rotating in a circular 
orbits.

The parameters governing the equilibrium configurations
are the mass ratio $q\equiv M_2/M_1<1$;  the scaled angular
velocity $\omega^2/G\rho$; and the separation between the
components center-of-masses, $d$. In Keplerian rotation,
the angular velocity, separation, and total mass are related 
through equation~(\ref{Kepler}). However, here we allow
departures from Keplerian rotation. When searching for 
equilibrium configurations, we treat the separation and 
angular velocity as two independent parameters, and solve
for the total mass of the system.

\begin{figure*}
\epsscale{1.0}
\plotone{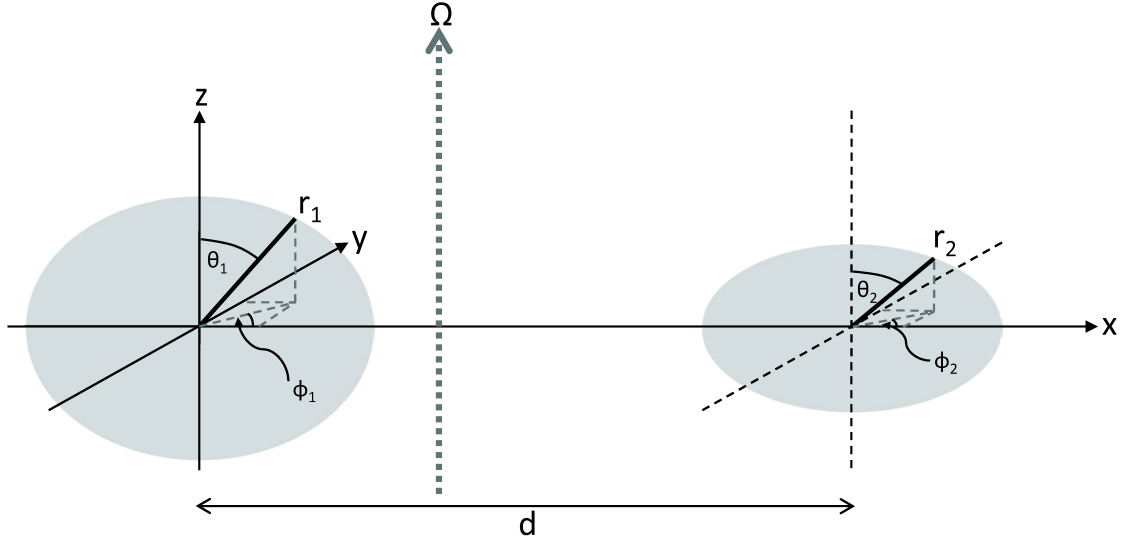}
\caption{A schematic representation of the coordinates used 
in the numerical solution.}
\label{schem-coord}
\end{figure*}

We first define a Cartesian coordinate system, whose origin 
is at the center of the more massive component, $M_1$.
We define the $\hat{x}$ direction to point toward the
lighter companion, $M_2$, which is centered at $x=d$. 
The common center of mass is located at $x=qd/(q+1)$.
We define the $\hat{z}$ direction to be parallel 
to the angular velocity.
Figure~\ref{schem-coord} illustrate the choice of
coordinates. 

We describe the surface of each body relative to its {\it own}
center of mass, using a spherical coordinate system.
The surface of each component is sampled along $N$ points with
given angular directions, specified by the azimuthal angle
$\phi$ and the polar angle $\theta$ defined in the usual way.
For example, in Figure~\ref{schem-coord},
\begin{equation}
\begin{array}{llll}
\left\{ 
\begin{array}{l}
x_1 = r_1 \sin(\theta_1) \cos(\phi_1)\\
y_1 = r_1 \sin(\theta_1) \sin(\phi_1)\\
z_1 = r_1 \cos(\theta_1)
\end{array}
\right. 
&&& 
\left\{ 
\begin{array}{l}
x_2 = d + r_2 \sin(\theta_2) \cos(\phi_2) \\
y_2 = r_2 \sin(\theta_2) \sin(\phi_2)\\
z_2 = r_2 \cos(\theta_2)
\end{array}
\right.
\end{array}\,.
\end{equation}
The surface mapping is performed by associating each direction 
$(\phi_i,\theta_i)$ with the distance, along this direction,
between the center of mass and the surface, $R_i$.

Each of the $N$ points corresponds to a small surface 
area surrounding it. We distribute the $N$ points evenly
in $\phi$ and in $\cos(\theta)$ to ensure that each point
covers an equal solid angle $\Delta\Omega$, as measured
from the center of the body. We assume that the 
equilibrium configurations are symmetric about the $x-y$
and $x-z$ planes. This assumption allows us to only sample
a quarter of a sphere when mapping the surface of each body.

Each surface patch corresponds to a ``mass-cone'' 
stretching from the patch to the center of the body.
For homogeneous bodies, the mass within the $i$'th ``cone'',
stretching from $r=0$ to the surface where $r=R_i$, 
is $M_i = \int_0^{R_i} \rho r^2 \Delta\Omega dr = 
\Delta\Omega \rho R_i^3/3$. The total mass in the body
is $M = \sum_i M_i = \Delta\Omega \rho / 3 \times \sum_i R_i^3$.

The potential at a point located on the surface of a 
body is the sum of the gravitational potentials induced
by the mass in all the cones in both components, and the
rotational potential about the common center of mass. 
Our goal is to solve for the values of $R_i$, for which the
total potential along the surface of each body is constant.
While the surface of each body must be equipotential, the
values of the potential may differ between the two bodies.

\subsection{Evaluating the Potential}
To compute the gravitational potential at some surface point,
we must first evaluate the distance between this point and 
a mass element $dm$ inducing the potential.
We can write the distance between two points that belong to 
$M_1$ or $M_2$, 
$(\phi_i,\theta_i,r_i)$ and $(\phi_j,\theta_j,r_j)$ as
\begin{equation} 
\label{dist}
\Delta r ^2 = r_i^2 + A r_i + B
\end{equation}
with
\begin{subequations}
\label{ABeq}
\begin{eqnarray}
\label{Aeq}
A = a r_j \pm 2 l \sin(\theta_i) \cos(\phi_i)\\
\label{Beq}
B = l^2 + r_j^2 \mp 2 l r_j \sin(\theta_j) \cos(\phi_j)
\end{eqnarray}
\end{subequations}
where
\begin{equation}
a = -2 \{ \sin(\theta_i) \sin(\theta_j) [\cos(\phi_i) \cos(\phi_j) + \sin(\phi_i) \sin(\phi_j)] + 
\cos(\theta_i) \cos(\theta_j)\}
\end{equation}
and
\begin{equation}
l = \left\{
\begin{array}{l}
d,\;\;\;i,j\;{\rm\;in\;different\;components} \\
0,\;\;\;i,j\;{\rm in\;same\;component}
\end{array} \right.\,.
\end{equation}
The plus (minus) sign in equation~\ref{Aeq} (\ref{Beq}) should be 
used if $j$ is in $M_1$ and $i$ is in $M_2$, and the minus (plus) 
sign should be used if $j$ is in $M_2$ and $i$ in $M_1$. 
If both points are in the same mass, $l=0$ such that
$A=a r_j$ and $B=0$.

Given this distance, the gravitational potential due to the
total mass in cone $i$ with extent $R_i$, at a point 
$(\phi_j,\theta_j,r_j)$ is
\begin{equation}
\begin{array}{lll}
V_i(r_j) =\Delta\Omega_i &\left\{ \left( \frac{R_i}{2} - \frac{3A}{4} \right) \; \sqrt{R_i^2 + A R_i + B} \right. \;\;
+ \frac{3A}{4}\sqrt{B}\;\; \\
&+ \left( \frac{3A^2}{8} - \frac{B}{2} \right) \ln\left( \frac{A}{2} + R_i + \sqrt{R_i^2 + A R_i + B} \right)\;\;\\
&- \left.\left( \frac{3A^2}{8} - \frac{B}{2} \right) \ln\left( \frac{A}{2} + \sqrt{B} \right)\,\right\},
\end{array}
\end{equation}
where $A$ and $B$ are the functions of $r_j$ specified
by equation~\ref{ABeq}, and $\Delta\Omega_i$ is the solid angle covered by cone $i$. 

For the rotational potential at a point $(\phi_j,\theta_j,r_j)$ 
we can write
\begin{equation}
V_{\rm rot} = \frac{1}{2} \omega^2 
\left\{ \begin{array}{lcl}
\left( \frac{q d}{q+1} \right)^2 + r_j^2 \sin(\theta_j)^2 - \frac{2 q d}{q+1} r_j \sin(\theta_j) \cos(\phi_j) &,& j $ in $ M_1\\
\left( \frac{d}{q+1} \right)^2   + r_j^2 \sin(\theta_j)^2 + \frac{2 d}{q+1}   r_j \sin(\theta_j) \cos(\phi_j) &,& j $ in $ M_2
\end{array} \right. \,.
\end{equation}

\subsection{The Newton-Raphson Scheme}
Consider a model with $N_1$ points sampling the surface of $M_1$,
and $N_2$ points sampling the surface of $M_2$.
The conditions of constant-potentials provides
$N_1+N_2$ equation. For points $R_j$ on the surface
of $M_k$ these equations can be written as,
\begin{equation}
\sum_{i \in M_1} V_i(R_j) + \sum_{i \in M_2} V_i(R_j) + V_{\rm rot}(R_j) - c_k =0\,.
\end{equation}
The variables in these equations are the $N_1+N_2$ values $R_i$, and
the values of the constant potentials on the surfaces of the two components,
$c_1$ on $M_1$, and $c_2$ on $M_2$.

Two additional equations are therefore required to solve the problem.
One of these equations is the constrain provided by the given mass ratio, $q$,
\begin{equation}
\label{massratio}
\frac{\Delta\Omega_2}{\Delta\Omega_1} \frac{\sum_{i \in M_1} R_i^3}{\sum_{j \in M_2} R_j^3} - q = 0\,,
\end{equation}
where $\Delta\Omega_1$ and $\Delta\Omega_2$ are 
the solid angles occupied by cones on $M_1$ and $M_2$, respectively.

The last equation is for the distance between the individual
center-of-masses of the two bodies,
\begin{equation}
\label{separation}
\frac{ \sum_{i \in M_1} \cos(\phi_i) \sin(\theta_i) R_i^4 / 4}{\sum_{i \in M_1} R_i^3 / 3} -
\frac{\sum_{j \in M_2} \cos(\phi_j) \sin(\theta_j) R_j^4 / 4} {\sum_{j \in M_2} R_j^3 / 3} 
= 0 \,.
\end{equation}

To summarize, the set of equation to be solved can be written as $\vec{F}=0$,
where $\vec{F}$ is the $N_1+N_2+2$ dimensional vector,
\begin{equation} \vec{F}^{(N1+N2+2)} \equiv 
\left\{ \begin{array}{l}
\vec{F}_{j\in M_1}^{(N_1)}\\
\vec{F}_{j\in M_2}^{(N_2)}\\
F_q^{(1)}\\
F_d^{(1)}
\end{array} \right\} 
=
\left\{
\begin{array}{l}
\sum_{i \in M_1,_2} V_i(Rj) + V_{\rm rot}(R_j) + c_1\\
\sum_{i \in M_1,_2} V_i(Rj) + V_{\rm rot}(R_j) + c_2\\
$Equation (\ref{massratio}) for the mass ratio$ \\
$Equation (\ref{separation}) for the separation$
\end{array} \right\}
= 0\,.
\end{equation}
The upper parentheses denote the size of each vector.

In the numerical procedure, we start with an initial guess
for the configurations, and iterate using a Newton-Raphson method
(but see~\ref{LS}),
correcting the values of our variables $\vec{x} = (\vec{R}_{j\in M_1},\vec{R}_{j\in M_2},c_1,c_2)$ by
$\vec{dx} = - \vec{F} \times {\bf J}^{-1}$, where ${\bf J}$ is the Jacobian derivatives matrix.

We compute the derivatives analytically.
We iterate the Newton-Raphson scheme until the solution has converged,
so that $\vec{F}=0$ to within some numerical threshold, and the
correction $\vec{dx}$ is small compared to $\vec{x}$.

\subsection{Symmetry}
As mentioned in Section~\ref{A1}, we assume that the
equilibrium configurations are symmetric about the $x-y$ and $x-z$
planes.
We therefore distribute $N$ points on the surface of a
$1/4$-sphere, between $\phi=0$ and $\pi$, and between 
$\cos(\theta)=1$ and $0$.

When computing the potential produced by the $i$'th cone
with orientation ($\mu_i=\cos(\theta_i), \phi_i)$, the contributions
of the three symmetric cones, at $(\mu_i, 2\pi-\phi_i)$,
$(-\mu_i, \phi_i)$ and $(-\mu_i, 2\pi-\phi_i)$
 must also be included.

The vector $\vec{F}$  therefore becomes,
\begin{equation}
\begin{array}{ll}
\vec{F} = \sum_i \left[
V_{R_i,\mu_i,\phi_i}(R_j) +
V_{R_i,\mu_i,2\pi-\phi_i}(R_j) +
V_{R_i,-\mu_i,\phi_i}(R_j) +
V_{R_i,-\mu_i,2\pi-\phi_i}(R_j)  \right] +\\
~~~~~~~
V_{\rm rot}(R_j) + c = 0\,,
\end{array}
\end{equation}
and the Jacobian matrix ${\bf J}$ must be corrected accordingly to
include the additional terms.

\subsection{A Least Squares-Newton Raphson Solution}
\label{LS}
We find that the numerical efficiency is much increased 
if instead of using equation~\ref{separation} for the distance between
the center-of-masses,  we use two equation,
\begin{subequations}
\begin{eqnarray}
\frac{ \sum_{i \in M_1} \cos(\phi_i) \sin(\theta_i) R_i^4 / 4}{\sum_{i \in M_1} R_i^3 / 3} = 0\\ 
\frac{ \sum_{i \in M_2} \cos(\phi_i) \sin(\theta_i) R_i^4 / 4}{\sum_{i \in M_2} R_i^3 / 3} = 0.
\end{eqnarray}
\end{subequations}
We therefore solve an over-constrained problem, with $N1+N2+3$ equations for
$N_1+N_2+2$ variables.
We substitute the Newton-Raphson matrix division 
$\vec{dx} = -\vec{F} \times {\bf J}^{-1}$,
with a so-called ``left-matrix division'' $\vec{dx} = -\vec{F}\, \backslash {\bf J}$,
which is the solution, in the least squares sense, to the over-constrained
system of equations, ${\bf J}\times \vec{dx} = -\vec{F}$.
We use the values of $\vec{dx}$ derived using this modified Newton-Raphson
scheme to correct the current solution, and iterate to convergence.

\subsection{Numerical Accuracy}

We consider the accuracy of our numerical method, as it applies to 
our best fit solution to the observed light curve of  Kuiper belt
binary 2001~QG$_{\rm 298}$. This model has $q=0.93$, $\Omega^2/G\rho=0.333$,
and $d=2.8956$. 

Given these parameters, we compare models constructed with various sampling
resolutions, namely $N=200, 400, 800, 1200$ and $1600$ points on a quarter-sphere. 
We begin by estimating the "true" equilibrium 
configurations, by extrapolating the functions $r_i(1/N)$ to $(1/N)=0$
for each direction $i=1\hdots N$.
For each $N$ we then compute the RMS difference between the numerical
solution and this "true" configuration, 
$\sqrt{\sum{(R_{i,{\rm sol}} - R_{i, {\rm true}})^2}/N}$, where
$R_{i, {\rm sol}}$ are the distances from the center-of-mass to the surface
in the numerical solution, and $R_{i, {\rm true}}$ are the distances from
the center-of-mass to the surface in the extrapolated, "true" solution.

Since each solution is given at different azimuthal and polar 
angles, we first interpolate the center-to-surface distances of the
high-resolution solutions onto the directions in which the lowest-resolution
solution is known. We use these interpolated versions of the high-resolution
solutions both to estimate the true configurations and to compute the RMS
differences.

\begin{figure}[!h]
\epsscale{0.5}
\plotone{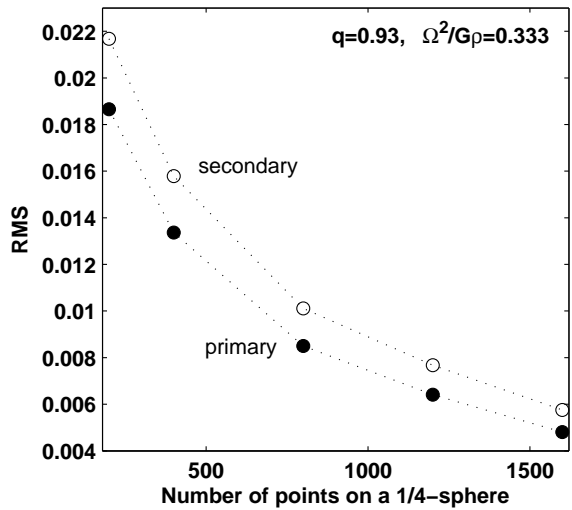}
\caption{An example for the RMS differences between ``true'' equilibrium configuration, and
numerical solutions at different resolutions (see text). The horizontal axis
shows the number of points on a quarter sphere. The filled symbols are for the more massive
primary, and the empty symbols are for the lower mass companion.}
\label{accuracy}
\end{figure}

Figure~\ref{accuracy} shows our results for the RMS difference
versus the number of points. The filled symbols are for the more massive
primary, and the empty symbols are for the lower mass companion. For example,
the RMS difference between the primary solution with $1600$ points and the 
"true" solution is $0.0048$. We note, that the interpolation process introduces
some errors that may increases this RMS. These values can be compared with
the  RMS difference between the numerical solution and the best fit ellipsoid
(see Section~4.1 and Figure~\ref{non-e}) which are $0.023$ and $0.027$ for the primary
and companion, respectively.

\section{Computing Light Curves}
\label{LCequations}

\subsection{Numerical Procedure}



To compute the observed luminosity, we tile the surface
of a given binary-configuration with triangles, typically two
triangles for each of the $N$ points used when computing
the equilibrium configuring. 
For the computations presented in sections~\ref{results} and~\ref{discussion}, 
which contain $\sim1600$-point on a quarter-sphere,
this corresponds to $\sim13,000$ triangular tiles on
each of the components.

We compute the $x,y,z$ coordinates of the vertices of
the triangular tiles. Given the orbital phase 
and inclination, we then rotate the coordinate system such 
that the positive $z$-direction points toward the observer.
Tiles with higher values of $z$ are closer to the 
observer, and may block tiles with lower values of $z$
from sight.

We now sort the tiles of both bodies according to their
mean $z$ values.
For each tile, we compute the area projected on the $x-y$
plane, $A^{x-y}_i$ , and the angle between the normal to the tile
surface and the Sun, $\vartheta_i$. 

We define a ``visibility'' function
for each tile, $V_i$, describing the fraction of the tile's
projected surface that is visible to the observer.
For each tile, we estimate the visibility function
by searching for other tiles which may block it from sight.
We start with the tile closest to the observer (highest $z$
value),
and search for possible overlaps between its $x-y$ position
and the $x-y$ positions of all the lower-$z$ tiles.
When an overlap exists, we reduce the visibility function
of the lower tile, by an amount corresponding to the
fraction of its area that overlaps with the higher tile,
$V_i = V_i - A^{\rm overlap}_{i,j}/A^{x-y}_i$.
We then repeat this procedure for every other tile,
correcting the visibility of lower tiles.

This provides an {\it overestimate} of the reduction in visibility,
since several high tiles may cover the {\it same} part of
a lower tile, resulting in a ``double-reduction'' of the
same covered area. This error remains small due to the simple shapes
considered here\footnote{We verified that we obtain the correct
projected area for the simple case of backscatter reflection from
two spherical or elliptical components.}.

The total observed light is given by 
$\sum_{i} A^{x-y}_i \times V_i$ for the case
of backscatter reflection, and by 
$\sum_{i} A^{x-y}_i \times V_i \times cos(\vartheta_i)$ 
for the case of diffuse (Lambertian) reflection.
To compute the light curve due to a given configuration viewed
at a given inclination, this process must be repeated for
a set of  orbital phases between $0$ and $2\pi$.

\subsection{Sanity Checks}
\label{sanity}

To verify the accuracy of our light-curve computations,
we compared our models against published results for
similar configurations. In particular we focused on previous
suggested models for QG$_{\rm 298}$. Two previous published
models exist, both using the Roche ellipsoidal solutions. 
The first is by Takahashi \& Ip (2004), and the second
by Lacerda \& Jewitt (2007).

\begin{figure}[!h]
\epsscale{0.5}
\plotone{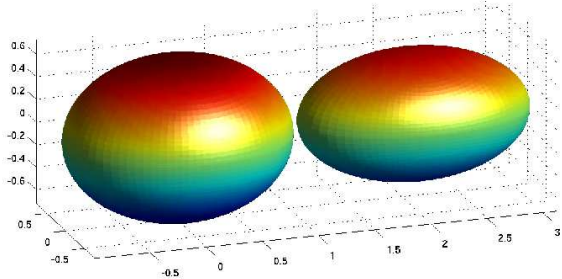}
\caption{The configuration found by Takahashi \& Ip~(2004) 
for QG$_{\rm 298}$.}
\label{TIconf}
\end{figure}

According to Takahashi \& Ip (2004), QG$_{\rm 298}$ consists
of two components with a mass ratio, $q=0.65$. The primary has 
$e_1=0.61$ and $e_2=0.69$, and a secondary has $e_1=0.79$ and
$e_2=0.83$. The separation is $2.1$ times the primary's semi-major
axis. The system is illustrated in Figure~\ref{TIconf}. 

Takahashi \& Ip find that their best fit indicates a 
$\sim70\%$ diffuse reflection plus $\sim30\%$ backscatter 
(uniform) reflection. For these parameters, they infer a
bulk density of $0.63\pm0.20$~g~cm$^{-3}$.

\begin{figure}[!h]
\epsscale{0.5}
\plotone{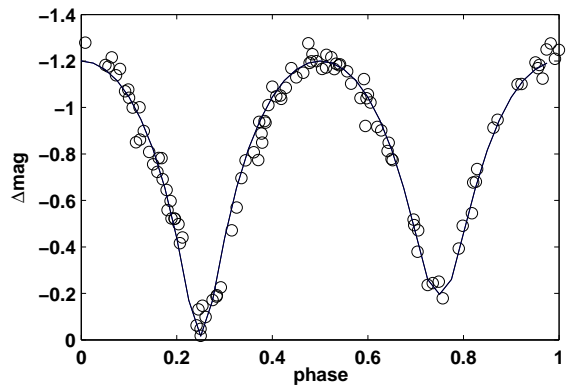}
\caption{The light curve that we compute for the parameters 
inferred by Takahashi \& Ip~(2004). Compare to their Figure~2.}
\label{TILC}
\end{figure}

Figure~\ref{TILC} shows the light curve that we compute given the
parameters inferred by Takahashi \& Ip, along with the observed
data points (left). A comparison of this light curve with that
presented in Figure~2 of Takahashi \& Ip (right), shows that our
results are nicely consistent.

\begin{figure}[!h]
\epsscale{0.5}
\plotone{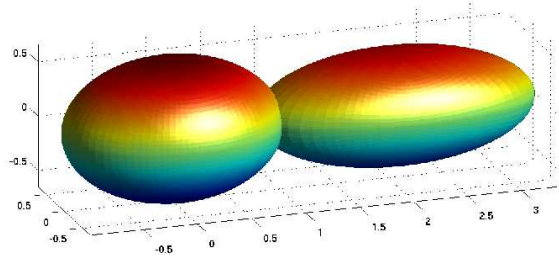}
\caption{The configuration found by Lacerda \& Jewitt~(2007) 
for QG$_{\rm 298}$.}
\label{LJconf}
\end{figure}

According to Lacerda \& Jewitt (2007), QG$_{\rm 298}$ is
a binary with a mass ratio $q=0.84$. Their primary has
$e_1=0.69$ and $e_2=0.76$, and their secondary $e_1=0.89$ and
$e_2=0.91$. The separation is $2.1$ times the primary semi-major axis.
This system is illustrated in Figure~\ref{LJconf}. Note that Lacerda \& Jewitt
accepted this configuration even though the two components
overlap. They argue that they chose to accept it, because of the inaccuracies 
imposed by the ellipsoidal and Keplerian nature of the Roche
approximation, and due to our poor understanding of the 
formation mechanisms of such binaries.

\begin{figure}[!h]
\epsscale{0.5}
\plotone{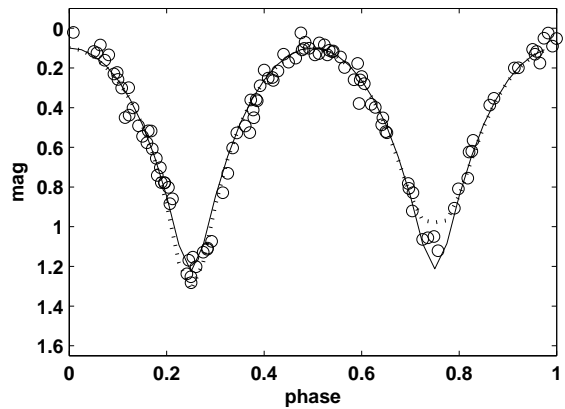}
\caption{The light curve that we compute for the parameters inferred
by Lacerda \& Jewitt~(2007). Compare to their Figure~9 (lower-left case).
Here the solid curves are for backscatter reflection, and the dashed
curves are for diffuse reflection.}
\label{LJLC}
\end{figure}

Lacerda \& Jewitt find that backscatter reflection best fits
the observed light curve. These parameters indicate a bulk
density of $0.590^{+0.143}_{-0.47}$~g~cm$^{-3}$.
In Figure~\ref{LJLC} we display the light curve that we compute
based on the parameters inferred by Lacerda \& Jewitt (2007).
Our light curves are again nicely consistent with those
computed by Lacerda \& Jewitt and shown in their Figure~9 (lower-left
panel).

\end{document}